\documentclass[man,12pt,floatsintext]{apa7}

\usepackage[american]{babel}
\usepackage{csquotes}
\usepackage[style=apa,sortcites=true,sorting=nyt,backend=biber]{biblatex}
\DeclareLanguageMapping{american}{american-apa}
\usepackage[T1]{fontenc}
\usepackage{mathptmx}
\usepackage{amsmath,amssymb,bm}
\usepackage{graphicx}
\usepackage{booktabs}
\usepackage{siunitx}
\usepackage{hyperref}
\usepackage{enumitem}
\usepackage{subcaption}
\usepackage{nameref}
\usepackage{threeparttable}
\usepackage{tabularx}
\usepackage{setspace}
\usepackage{array}
\usepackage{multirow}
\usepackage[table]{xcolor}
\usepackage{placeins}
\usepackage{tikz}
\usetikzlibrary{arrows.meta,positioning,fit,calc}

\definecolor{rulegray}{gray}{0.82}
\definecolor{bandgray}{gray}{0.97}
\newcolumntype{L}[1]{>{\raggedright\arraybackslash}m{#1}}
\newcolumntype{C}[1]{>{\centering\arraybackslash}m{#1}}

\title{Slow Context, Fast Symptoms: Multiscale Temporal Dynamics and Context-Induced Coupling in Psychological Systems}
\shorttitle{Slow--Fast Psychological Dynamics}

\author{%
Kyuri Park\textsuperscript{*1},
Denny Borsboom\textsuperscript{3},
Mike Lees\textsuperscript{1,2},
Lourens Waldorp\textsuperscript{3},
Johan Bollen\textsuperscript{1},
V\'{i}tor V.\ Vasconcelos\textsuperscript{1,2}
}
\affiliation{%
\textsuperscript{1} University of Amsterdam\mbox{,} Informatics Institute\mbox{,} Computational Science Lab\\
\textsuperscript{2} University of Amsterdam\mbox{,} Institute for Advanced Study\\
\textsuperscript{3} University of Amsterdam\mbox{,} Department of Psychology
}
\authornote{%
\textbf{Author contact information.}
Kyuri Park (\href{mailto:k.park@uva.nl}{k.park@uva.nl});
V\'{i}tor V.\ Vasconcelos (\href{mailto:v.v.vasconcelos@uva.nl}{v.v.vasconcelos@uva.nl});
Mike Lees (\href{mailto:m.h.lees@uva.nl}{m.h.lees@uva.nl});
Denny Borsboom (\href{mailto:d.borsboom@uva.nl}{d.borsboom@uva.nl});
Lourens Waldorp (\href{mailto:l.j.waldorp@uva.nl}{l.j.waldorp@uva.nl});
Johan Bollen (\href{mailto:j.l.t.m.bollen@uva.nl}{j.l.t.m.bollen@uva.nl}).\\[0.5em]
\textbf{Correspondence.}
*Correspondence concerning this article should be addressed to Kyuri Park.
}

\abstract{
\begingroup
\setstretch{1.3}\selectfont
Changes in psychological states can occur over short timescales, while the context that shapes those changes often evolves more slowly. Symptoms, affect, and attention can shift quickly, whereas stress exposure, social isolation, financial strain, biological vulnerability, and environmental context may persist over longer periods. This difference in timescale matters because slow context can change how easily short-term states become active, how long they persist, and how easily they recover.
We develop this argument through the case of psychopathology symptom networks. Network models often focus on symptom--symptom coupling, but symptoms also unfold within slow context. Context can be understood as changing symptom activation tendencies, but the feedback between the two is less often made explicit. Slow context can shape fast symptom activation, and sustained symptom activation can, in turn, feed back into slow context.
We formalize this coupling in a computational slow--fast model. Simulations show that contextual input can shift symptom activation without changing symptom--symptom coupling. Perturbations to context can produce temporary symptom increases followed by recovery. Symptom-to-context feedback can slow recovery and, when stronger, produce initial-state dependence. When between-person contextual variation is omitted from an estimated symptom network, estimated coupling also increases, and additional coupling appears among symptom pairs that are uncoupled in the data-generating model.
The resulting framework connects fast psychological states with the slower conditions in which they unfold. Although developed through symptom networks, it applies more generally to psychological systems in which processes operating at different timescales interact and shape one another.
\par
\endgroup
}


\begin{document}
\maketitle


Changes in psychological states can occur over short timescales, while the
context that shapes those changes often evolves more slowly. Symptoms, affect, and
attention can shift over short timescales, while stress exposure, social isolation, financial strain, biological vulnerability, and environmental context often change more slowly.
Slow context does not simply sit outside psychological dynamics. It can change
how easily faster states become active, how long they persist, and how easily
they recover.

Psychopathology symptom networks offer a useful case for studying this
slow--fast relation. Network approaches start from the idea that symptoms may do more than co-occur: they may also influence one another. Insomnia may contribute to fatigue, fatigue may reduce concentration, and concentration problems may worsen feelings of failure.
In this view, symptoms are not only indicators of an underlying disorder. They
are parts of a system whose dynamics can become self-sustaining \parencite{borsboom2013network, Borsboom2017, fried2020systems,
robinaugh2020network}. Network theories have also used ideas from dynamical-systems theory to describe how symptom activation may persist after an initial trigger has disappeared, and how symptom systems may sometimes show abrupt changes, hysteresis-like dynamics, or early warning signals
\parencite{cramer2016major, van2014critical, helmich2021early,
Dablanderetal2023, scheffer2009early, park2026feedback}.

At the same time, symptoms do not unfold outside the context of a person's
life. Stressful events, social isolation, financial insecurity, sleep
disruption, inflammation, and other contextual or biological processes can make symptoms easier or harder to activate. This idea is consistent with broader vulnerability--stress and resilience perspectives, in which relatively stable or slowly changing context shapes how people respond to later events \parencite{ingram2005vulnerability, McEwen1998, masten2021resilience, lunansky2020personality}. In network terms, slow context can be understood as part of the external input (often called the field) of the symptom system. Here, external means external to the modeled symptom network, not necessarily external to the person or the body. We use slow context specifically for processes that change more slowly than the symptom states of interest and modulate their propensity for activation. These may include social and economic conditions, such as poverty or isolation \parencite{elsenburg2026precariousness, park2026precariousness}, as well as biological or developmental processes.
This distinction matters because the same symptom network may behave differently depending on the slow context in which it is embedded. A perturbation may fade when symptoms are difficult to activate but persist when slow context makes activation easier or recovery slower.

Studies that repeatedly measure people in daily life have made short-term psychological fluctuations increasingly visible, especially by showing how symptoms, affect, and behaviour change together within individuals over time \parencite{bringmann2013network, myin2009experience, wichers2014dynamic, Epskampetal2018, neubauer2020studying}. This paper asks what happens when these short-term dynamics are embedded in a slower contextual process. We do not assume that slow context necessarily affects symptoms, or that symptom activation necessarily feeds back into context. Rather, the model allows these influences to vary in strength or to be absent.

The issue arises first at the level of individual dynamics. If an unmeasured slow process influences several symptoms at once, changes in that process can induce covariation among symptoms even when their underlying symptom–symptom coupling has not changed. An estimated symptom network may then partly reflect shared contextual input rather than interactions among symptoms themselves. 
Between-group comparisons introduce an additional version of the same problem. If groups systematically differ in their slow contextual conditions, separately estimated networks may differ even when the underlying symptom--symptom coupling is identical. This concern is particularly relevant for observationally defined groups that differ in illness course, treatment stage, exposure history, or related characteristics \parencite[e.g.,][]{van2015association, beard2016network, peel2021comparison}. Such studies may compare global strength, edge weights, centrality, or moderated network parameters \parencite{haslbeck2022estimating, van2023comparing}. Differences in these quantities may then be interpreted as evidence for differences in the organization or strength of symptom--symptom relations.


But this interpretation is not forced by the data. Throughout this paper, we use symptom--symptom coupling in the theoretical sense: activation of one symptom changes the activation tendency of another symptom in the data-generating process. Estimated edges or interaction parameters are statistical quantities. They may be interpreted as evidence for such coupling only under additional assumptions about the data-generating process. A difference in estimated connectivity may therefore reflect true differences in symptom--symptom coupling, but it may also reflect differences in slow context. This ambiguity is related to existing equivalence results in network psychometrics. Associations among binary symptoms can often be represented in more than one way: as direct symptom--symptom interactions, as dependence induced by shared latent or common causes, or as item-response structure
\parencite{kruis2016three, marsman2018introduction, vanbork2021latent}. These
results imply that observed symptom associations do not uniquely identify
symptom--symptom coupling. A related perspective is that some apparent network
differences may reflect changes in activation tendencies or contextual input
rather than changes in symptom--symptom coupling itself \parencite{park2026rethinking}.

The contribution of the present paper is therefore not the claim that external fields can affect symptom activation. That idea is already native to network theory and to Ising-type formulations of binary symptom data \parencite{van2014new, marsman2018introduction}. Our focus is what changes when contextual processes and symptom dynamics evolve on different, but coupled, timescales. Slow context can gradually shift the activation landscape of the faster symptom system, while sustained symptom activation can, in turn, feed back into that context. This time-scale separation can produce delayed recovery, persistence and history dependence even when the underlying symptom--symptom coupling remains unchanged. It also creates an inferential problem: observed symptom relations can depend on where and when the system is observed within the slower contextual process.

Related methodological work makes a similar point from a statistical direction:
estimated pairwise relations do not always recover direct interactions in the
data-generating process. For Ising models and related binary graphical models,
nodewise regularized logistic regression provides one foundation for estimating
graph structure under appropriate conditions \parencite{ravikumar2010high}.
More recent work extends pseudo-likelihood approaches to models that include
both pairwise interactions and covariate-dependent external fields
\parencite{mukherjee2024logistic}. Work on reconstructability in dynamical
networks likewise shows that pairwise statistical dependence need not recover
direct interactions when common-cause, relay, or topological effects contribute
to observed associations, even though weak-coupling regimes can improve
reconstructability \parencite{lunsmann2017transition}. These lines of work are
methodologically adjacent to the present paper. Here, we use a known
data-generating system to ask what happens when slow context is part of the
process but not directly observed in the estimated symptom network.

The present paper develops a computational slow--fast model of symptom dynamics.
The model builds directly on earlier slow--fast thinking in psychopathology,
especially work linking slower personality and resilience processes to faster
symptom networks \parencite{lunansky2020personality}. The model embeds a fast
\(0/1\) symptom network within a slow contextual field. The fast layer
represents symptom activation and symptom--symptom coupling. The slow layer
represents context that can shift symptom activation, recover slowly, fluctuate, be perturbed, or be changed by sustained symptom activation. We use simulation as a theory-construction tool \parencite{haslbeck2022modeling, park2026causalnet}: the aim is to make clear what follows from this slow--fast structure, and what does not, under a known data-generating process.

We use the model in four ways. First, we vary the slow contextual field while holding symptom--symptom coupling fixed to examine how context changes overall symptom activation. We then perturb the slow contextual field and examine how symptom activation increases and recovers. Next, we introduce symptom-to-context feedback to study whether it alters recovery and allows earlier symptom activation to shape later dynamics through changes in context. Finally, we test how well the underlying symptom--symptom coupling is recovered when the slow contextual process is present in the data-generating system but omitted from the estimated network.

Across all four analyses, the same slow--fast model is used to examine different consequences of context acting on symptom dynamics. The final analysis also points to a practical implication: when slower contextual processes are plausible, measuring them alongside symptoms may help distinguish symptom--symptom coupling from associations induced by shared contextual input. The approach provides a computational framework for studying psychological dynamics across timescales without treating fast symptom dynamics and slow context as isolated parts of the system.

\section{A slow--fast model of symptom dynamics}

We introduce the model in stages. We begin with fast probabilistic dynamics for
binary symptom states. We then add a slow contextual field \(P_t\), which shifts
symptom activation without changing symptom--symptom coupling. Finally, we allow
\(P_t\) to evolve over time, be perturbed, and receive feedback from sustained
symptom activation.

\begin{figure}[htbp]
\centering
\includegraphics[width=0.95\textwidth]{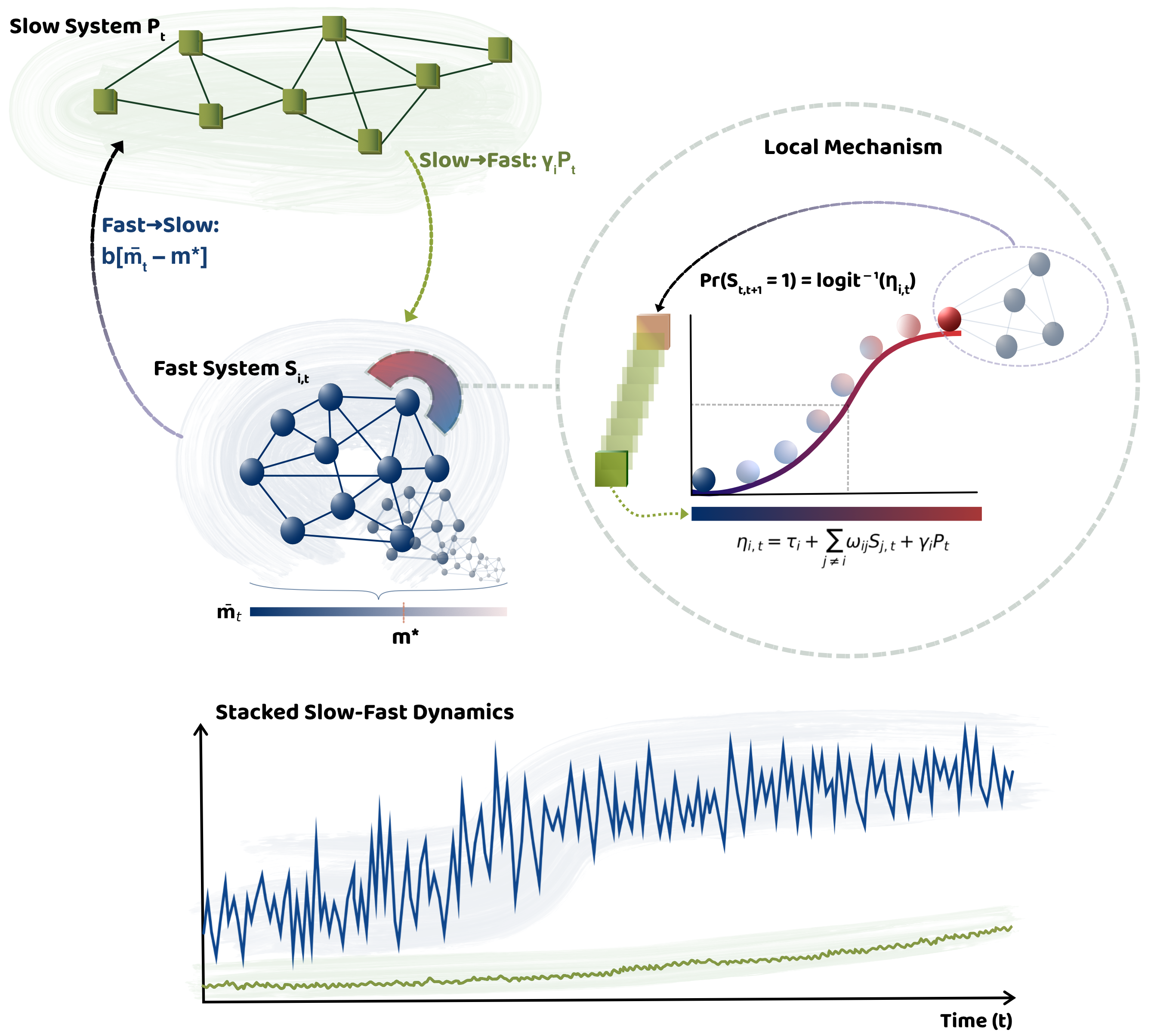}
\caption{\textbf{Conceptual schematic of the slow--fast symptom model.}
The fast layer consists of binary symptom states \(S_{i,t}\in\{0,1\}\) that
activate and deactivate on a short timescale. As in standard Ising symptom-network models, symptoms interact through the coupling matrix \(\omega\). 
The extension introduced here is that this fast symptom system is embedded in a slower contextual process \(P_t\), which evolves more gradually and shifts symptom activation through the additive input \(\gamma_iP_t\).
The green contextual layer is drawn schematically as structured contextual conditions, but these conditions are summarized in the present model by the scalar field \(P_t\), not by an explicitly modeled second network. The local activation rule combines a symptom-specific baseline tendency \(\tau_i\), symptom--symptom input \(\sum_{j\neq i}\omega_{ij}S_{j,t}\), and slow contextual input \(\gamma_iP_t\) into \(\eta_{i,t}\), which determines the probability that symptom \(i\) becomes active. The model also allows feedback from the fast layer to the slow layer: sustained mean symptom activation, summarized by \(\bar m_t\), enters the slow-field update through \(b(\bar m_t-m^\star)\). When recent symptom activation is above \(m^\star\), this term pushes the slow field upward; when it is below \(m^\star\), it pushes the slow field downward. Acute perturbations enter the slow field through\(\xi_t\).}
\label{fig:slow-fast-illust}
\end{figure}

Figure~\ref{fig:slow-fast-illust} gives the basic structure of the model before
we define each equation. A summary of the main model components is given at the
end of this section in Table~\ref{tab:model_components}.

\subsection{Fast symptom network}

We first define the fast layer as a probabilistic network model (PNM) for binary symptom states. In the binary case, this formulation is closely related to Ising models used in network psychometrics
\parencite{van2014new, marsman2018introduction}. Let \(S_i\in\{0,1\}\) denote
whether symptom \(i\) is inactive or active. We write the local input to symptom \(i\) as

\begin{equation}
\eta_i
=
\tau_i+\sum_{j\neq i}\omega_{ij}S_j,
\label{eq:fast_input}
\end{equation}

so that its activation probability is

\begin{equation}
\Pr(S_i^{\mathrm{new}}=1\mid \mathbf S_{-i})
=
\operatorname{logit}^{-1}(\eta_i).
\label{eq:fast_activation}
\end{equation}

At the level of this local update, the formulation has the same form as a
logistic generalized linear model: \(\tau_i\) acts as the symptom-specific
intercept, while the states of the other symptoms enter as predictors with
coefficients \(\omega_{ij}\). Here, \(\omega_{ij}\) represents the pairwise
coupling between symptoms \(i\) and \(j\). Thus, \(\tau_i\) gives the activation tendency of symptom \(i\) when it receives no input from the rest of the network, while active symptoms with nonzero couplings shift this tendency through the summed interaction term.

In the simulations, one fast step consists of a random-order sweep through the
symptoms, updating each symptom once conditional on the current states of the
others. We write \(S_i^{\mathrm{new}}\) to emphasize that the update is local to symptom \(i\), rather than a synchronous update of all symptoms at once.

\subsection{Adding slow context}

The slow--fast extension adds a contextual state \(P_t\) that evolves more
slowly than symptom activation. Rather than changing the symptom--symptom
couplings, slow context enters as an additional term in the local input:

\begin{equation}
\eta_{i,t}
=
\tau_i+\sum_{j\neq i}\omega_{ij}S_{j,t}+\gamma_iP_t .
\label{eq:slow_fast_input}
\end{equation}

The corresponding activation probability is

\begin{equation}
\Pr(S_i^{\mathrm{new}}=1\mid \mathbf S_{-i,t},P_t)
=
\operatorname{logit}^{-1}(\eta_{i,t}) .
\label{eq:slow_fast_activation}
\end{equation}

Here, \(\gamma_i\) determines how strongly the slow contextual state shifts the activation tendency of symptom \(i\). When \(\gamma_i=0\), symptom \(i\) is unaffected by \(P_t\); larger positive values make its activation increasingly sensitive to changes in slow context. We use the \(\eta_{i,t}\) notation in Figure~\ref{fig:slow-fast-illust} to summarize the local activation rule visually.

The contextual state \(P_t\) summarizes conditions outside the modeled symptom
network that evolve on a slower timescale and can change how easily symptoms
become active. Here, contextual means external to the symptom network, not
necessarily outside the person or the body. It may represent social or
environmental conditions, such as financial strain or housing insecurity, as well as slower biological or developmental processes. Figure~\ref{fig:slow-fast-illust} depicts these conditions schematically as a structured contextual layer. In the present simulations, however, this layer is not modeled as a second network but is summarized by the single scalar state \(P_t\).

\subsection{Slow contextual dynamics}

The previous subsection introduced \(P_t\) as an input to the symptom network.
We now specify how this slow contextual field changes over time. The two
directions of coupling are shown in Figure~\ref{fig:slow-fast-illust}. The
green slow-to-fast pathway shows how \(P_t\) shifts symptom activation through
\(\gamma_iP_t\). The blue fast-to-slow pathway shows how sustained symptom
activation, summarized by \(\bar m_t\), can feed back into \(P_t\). Acute
perturbations enter the same slow field through the jump term \(\xi_t\).

The slow-field update is
\begin{equation}
P_{t+dt}
=
P_t
+
\left[
-\kappa(P_t-P_{\mathrm{base}})
+
b(\bar m_t-m^\star)
\right]dt
+
\sigma_P\sqrt{dt}\,\epsilon_t
+
\xi_t,
\qquad
\epsilon_t\sim\mathcal N(0,1).
\label{eq:context_update}
\end{equation}

Each term corresponds to a part of Figure~\ref{fig:slow-fast-illust}. The first
term, \(P_t\), is the current level of the slow contextual field. The term
\(-\kappa(P_t-P_{\mathrm{base}})dt\) pulls the slow field back toward its
baseline \(P_{\mathrm{base}}\). The parameter \(\kappa\) controls how quickly
this return occurs. The noise term
\(\sigma_P\sqrt{dt}\,\epsilon_t\) represents small background fluctuations in
the slow field, where \(\epsilon_t\) is standard normal noise.

The jump term \(\xi_t\) represents occasional larger perturbations to the slow
field. Conceptually, these are identifiable events with specific timing and
magnitude, such as conflict, loss, financial setback, health problems,
relationship change, birth, or death
\parencite{cohen2019ten, dohrenwend2006inventorying}. This is why \(\xi_t\) is added directly to \(P_t\): when a perturbation occurs,
it produces a discrete shift in the contextual state. In the recovery simulations reported here, we use a single deterministic perturbation, \(\xi_t=1\) at perturbation onset and \(\xi_t=0\) otherwise; Appendix~\ref{app:model-details} also describes a more general stochastic event formulation.

The feedback term,
\[
b(\bar m_t-m^\star)dt,
\]
corresponds to the fast-to-slow pathway in Figure~\ref{fig:slow-fast-illust}.
It allows sustained symptom activation to feed back into the slow field. Here,

\begin{equation}
m_t
=
\frac{1}{N}\sum_{i=1}^{N}S_{i,t},
\label{eq:mean_activation}
\end{equation}

is the instantaneous mean symptom activation. The quantity \(\bar m_t\) is a smoothed version of \(m_t\): it summarizes recent symptom activation rather than only the current moment. This smoothing is what gives the feedback pathway its
slow timescale: the slow field responds to sustained activation, not to every momentary symptom fluctuation. This matters because
the slow field is meant to summarize slower contextual conditions. A single bad moment should not immediately change these conditions, but symptom activation that persists
over time may affect work, relationships, sleep, stress exposure, or recovery.
The smoothing process used to compute \(\bar m_t\) is described in Appendix~\ref{app:model-details}.

The reference level \(m^\star\) determines the level of recent symptom
activation around which feedback changes direction, and \(b\) controls the
strength of this feedback. When \(\bar m_t>m^\star\), sustained symptom
activation pushes the slow field upward. When \(\bar m_t<m^\star\), the
feedback term pushes the slow field downward. This captures a feedback relation
between symptoms and slow context, consistent with stress-generation ideas:
symptoms may not only respond to stressful conditions, but also increase later
exposure to stressors or make existing conditions harder to recover from
\parencite{hammen2006stress, liu2010stress, rnic2023vicious, park2026precariousness}.

In the simulations below, we isolate different parts of the model by setting
some terms to zero. Setting \(b=0\) removes symptom-to-context feedback, whereas
setting \(\xi_t=0\) removes acute perturbations. This lets us distinguish the
slow-to-fast pathway, where slow context shapes symptom activation, from the
fast-to-slow pathway, where sustained symptom activation feeds back into the
slow field.

Table~\ref{tab:model_components} summarizes the main components of the model.
The table is meant as a reference for the equations above: the fast layer
contains symptom states, symptom-specific activation tendencies, and
symptom--symptom coupling; the slow layer contains the contextual field, its
baseline dynamics, perturbations, and feedback from sustained symptom activation.

\definecolor{bandgray}{gray}{0.97}
\definecolor{rulegray}{gray}{0.80}
\newcolumntype{L}[1]{>{\raggedright\arraybackslash}m{#1}}

\begin{table}[htbp]
\centering
\begin{threeparttable}
\caption{Main components of the slow--fast model.}
\label{tab:model_components}
\footnotesize
\setlength{\tabcolsep}{6pt}
\renewcommand{\arraystretch}{1.15}
\begin{tabularx}{\linewidth}{@{} L{3.2cm} L{2.7cm} X @{}}
\toprule
\rowcolor{bandgray}
\multicolumn{3}{c}{\textit{Fast symptom layer}} \\
\arrayrulecolor{rulegray}\specialrule{0.4pt}{0pt}{0pt}
\textbf{Component} & \textbf{Symbol} & \textbf{Interpretation} \\
\arrayrulecolor{rulegray}\specialrule{0.4pt}{0pt}{0pt}
Symptom state
  & \(S_{i,t}\)
  & Whether symptom \(i\) is inactive or active at time \(t\). \\
Baseline activation
  & \(\tau_i\)
  & Baseline log-odds of activation when symptom \(i\) receives no input from the rest of the network. \\
Symptom coupling
  & \(\omega_{ij}\)
  & Pairwise coupling between symptoms \(i\) and \(j\); an active symptom \(j\) contributes \(\omega_{ij}\) to symptom \(i\)'s local input. \\
Local input
  & \(\eta_{i,t}\)
  & Total input determining symptom \(i\)'s activation probability. \\
Slow-context loading
  & \(\gamma_i\)
  & Sensitivity of symptom \(i\)'s activation tendency to the slow contextual state \(P_t\). \\
Mean symptom activation
  & \(m_t\)
  & Proportion of symptoms active at time \(t\). \\
Recent symptom activation
  & \(\bar m_t\)
  & Smoothed recent mean symptom activation used in the feedback term. \\
Smoothing rate
  & \(\lambda_m\)
  & Weight given to current mean symptom activation when updating
    \(\bar m_t\); smaller values imply longer memory. \\
\midrule
\rowcolor{bandgray}
\multicolumn{3}{c}{\textit{Slow contextual process}} \\
\arrayrulecolor{rulegray}\specialrule{0.4pt}{0pt}{0pt}
\textbf{Component} & \textbf{Symbol} & \textbf{Interpretation} \\
\arrayrulecolor{rulegray}\specialrule{0.4pt}{0pt}{0pt}
Contextual state
  & \(P_t\)
  & Slow contextual state that evolves over time and shifts symptom activation. \\
Contextual baseline
  & \(P_{\mathrm{base}}\)
  & Level toward which \(P_t\) tends to return. \\
Return-to-baseline rate
  & \(\kappa\)
  & Speed with which \(P_t\) returns toward \(P_{\mathrm{base}}\). \\
Background fluctuation
  & \(\sigma_P\sqrt{dt}\epsilon_t\)
  & Small stochastic fluctuation in the slow contextual state. \\
Perturbation
  & \(\xi_t\)
  & Occasional larger jump in the slow state, such as an acute stressor. \\
Feedback reference
  & \(m^\star\)
  & Reference level around which symptom-to-context feedback changes direction. \\
Feedback strength
  & \(b\)
  & Strength of symptom-to-context feedback. \\
Step size
  & \(dt\)
  & Time step used for the slow contextual process update. \\
\arrayrulecolor{black}\bottomrule
\end{tabularx}
\vspace{2mm}
\begin{tablenotes}[flushleft]
\footnotesize
\item \textit{Note.} The table summarizes the components used in the model
equations. The green contextual layer in Figure~\ref{fig:slow-fast-illust} is
schematic; in the simulations, slow context is represented by the scalar contextual state \(P_t\).
\end{tablenotes}
\end{threeparttable}
\end{table}





\section{Simulation overview}

The simulations use the model as a controlled theoretical tool. They are not
intended to calibrate a specific clinical population. Instead, they ask what
kinds of patterns can arise when fast symptom dynamics are embedded in a slower
contextual field. Across simulations, the symptom--symptom coupling matrix is
held fixed, while the role of slow context is varied. We begin with differences
in the level of the slow contextual field, then add an acute perturbation, then
allow sustained symptom activation to feed back into the slow field, and finally
ask how the same slow--fast process appears when the slow field is omitted from
an estimated symptom network.

\subsection{Simulation 1: Slow-context level}

Simulation 1 provides a reference case for the slow-to-fast pathway. It asks
how a fixed contextual level changes symptom activation when the fast symptom
network itself is unchanged. Because \(P\) enters each symptom's local input
through \(\gamma_iP\), a shift in activation is expected from the model
definition. The purpose of this first simulation is therefore to establish the
baseline slow-to-fast effect against which the dynamic simulations that follow
can be interpreted.

\paragraph{Design.}
The fast layer contained \(N=9\) binary symptoms, corresponding to the nine
symptom items of the PHQ-9 \parencite{kroenke2001phq}. To isolate the effect
of slow context on symptom activation, we held the symptom-specific baseline
parameters \(\tau_i\), symptom--symptom couplings \(\omega_{ij}\), and
slow-context loadings \(\gamma_i\) fixed across conditions and varied only the
contextual state:
\[
P\in\{-0.6,0,0.6\}.
\]
The symptom-specific parameters and contextual levels were specified for
theoretical illustration rather than estimated from empirical data. The three
values of \(P\) represent low, reference, and high slow-context conditions on
the simulation scale. All slow-context loadings were positive and
symptom-specific, ranging from \(0.60\) to \(1.00\), so increasing \(P\)
raises the activation tendency of each symptom, although by different amounts.
The selected values of \(P\) produce graded changes in activation rather than
floor or ceiling saturation.
Within each condition, \(P\) remained fixed throughout the simulation.
Symptom states were updated according to
Eqs.~\eqref{eq:slow_fast_input}--\eqref{eq:slow_fast_activation}, with all
other model parameters held identical across conditions.

We simulated 200 independent chains per condition, with each symptom
initialized independently from a Bernoulli distribution with probability
\(0.5\). Each chain ran for 400 sweeps. The first 200 sweeps were discarded
as burn-in, and the remaining 200 sweeps were used to summarize symptom
activation. A post-burn-in stability check showed negligible remaining drift
relative to the difference between contextual conditions
(Appendix~\ref{app:sim1-stability}). We summarized each condition by the
distribution and mean number of active symptoms and by the proportion of
post-burn-in states with at least five active symptoms. Table~\ref{tab:sim1_parameters} summarizes the simulation
design. The exact symptom-specific parameter values are reported in
Appendix~\ref{app:main-parameters}.

\begin{table}[htbp]
\centering
\caption{Simulation 1 design.}
\label{tab:sim1_parameters}
\footnotesize
\setlength{\tabcolsep}{6pt}
\renewcommand{\arraystretch}{1.12}
\begin{tabularx}{\linewidth}{@{} L{3.2cm} X X @{}}
\toprule
\textbf{Parameter} & \textbf{Meaning} & \textbf{Value in Simulation 1} \\
\midrule
\(N\) & Number of symptoms & 9 \\
\(P\) & Fixed contextual state & \(-0.6, 0, 0.6\) \\
\(\tau_i\) & Baseline log-odds of activation
  & Symptom-specific; \(-2.70\) to \(-0.60\)
    (Table~\ref{tab:main_symptom_parameters}) \\
\(\omega_{ij}\) & Symptom coupling
  & 10 positive nonzero couplings; \(0.20\)--\(0.45\)
    (Table~\ref{tab:main_couplings}) \\
\(\gamma_i\) & Slow-context loading
  & Symptom-specific; \(0.60\)--\(1.00\)
    (Table~\ref{tab:main_symptom_parameters}) \\
Initial state & Initial symptom activation
  & Bernoulli\((0.5)\), independently by symptom \\
Chains & Independent chains per condition & 200 \\
Total sweeps per chain & Full simulated trajectory length & 400 \\
Burn-in & Initial discarded sweeps & First 200 sweeps \\
\bottomrule
\end{tabularx}
\end{table}

\begin{figure}[ht]
\centering
\includegraphics[width=\textwidth]{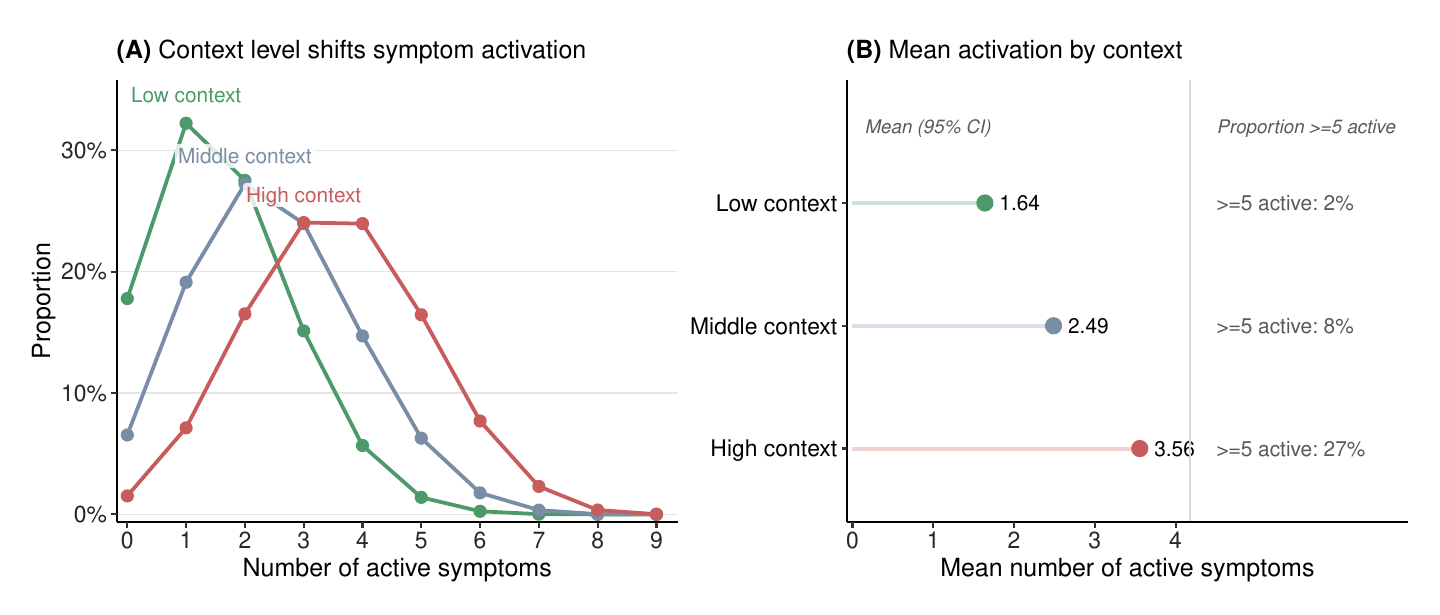}
\caption{\textbf{Slow contextual level shifts overall symptom activation under
fixed symptom coupling.}
(A) The distribution of the number of active symptoms shifts toward higher
activation as the slow contextual level \(P\) increases. (B) The mean number of
active symptoms and the proportion of post-burn-in states with five or more
active symptoms both increase across slow-context conditions.}
\label{fig:context-baseline}
\end{figure}

\paragraph{Results.}
Figure~\ref{fig:context-baseline} shows the expected shift in symptom
activation as the contextual state \(P\) increases while
symptom--symptom coupling remains fixed. Panel A shows the distribution of
the number of active symptoms across post-burn-in states. The distribution
moves progressively toward higher activation from \(P=-0.6\) to \(P=0\) to
\(P=0.6\).
Panel B summarizes the same pattern. The average number of active symptoms was
1.64 in the low-context condition (\(P=-0.6\)), 2.49 in the reference
condition (\(P=0\)), and 3.56 in the high-context condition (\(P=0.6\)).
The upper tail of the distribution shifted similarly: the proportion of
post-burn-in states with five or more active symptoms increased from \(0.017\)
to \(0.084\) and \(0.268\), respectively.

This graded shift is the expected consequence of the slow-to-fast term
\(\gamma_iP\). Because all \(\gamma_i\) are positive, sufficiently large
positive values of \(P\) would eventually dominate the local input and drive
the system toward ceiling activation. The contextual levels used here remain
well below that regime and instead produce a graded shift in symptom
activation.

Simulation 1 therefore serves primarily as a reference case: it shows how a
common contextual input can change the state occupied by an otherwise
unchanged symptom network. The slow--fast dynamics become consequential in
the simulations that follow, where \(P_t\) evolves on its own timescale,
recovers after perturbation, and can receive feedback from sustained symptom
activation.

\subsection{Simulation 2: Perturbing slow context}

Simulation 2 asks how symptom activation changes when the contextual state
\(P_t\) is temporarily pushed upward and then allowed to recover. Unlike
Simulation 1, \(P_t\) now evolves over time, making the separation between the
fast symptom dynamics and the slower contextual process explicit. The
perturbation is applied to \(P_t\), not directly to the symptoms; symptoms
respond only through the slow-to-fast input \(\gamma_iP_t\).

\paragraph{Design.}
Simulation 2 used the same symptom-specific baselines \(\tau_i\),
symptom--symptom couplings \(\omega_{ij}\), and slow-context loadings
\(\gamma_i\) as Simulation 1. The symptom-specific loadings therefore remained
between \(0.60\) and \(1.00\). We allowed \(P_t\) to evolve according to
Eq.~\eqref{eq:context_update}, but set \(b=0\) so that symptoms did not feed
back into the contextual process. This isolates the slow-to-fast pathway.

We set \(P_{\mathrm{base}}=0\). After 200 pre-perturbation steps, we introduced a single positive perturbation of size \(\xi_t=1\). We then simulated the system for an additional 750 steps. The simulation uses model time units rather than calibrated real-world time: \(dt\) and the number of steps specify the relative separation between the fast symptom dynamics and the slower contextual process, not a direct mapping onto hours, days, or months. At each step, the fast symptom network was updated conditional on the current value of \(P_t\). Table~\ref{tab:sim2_parameters} summarizes the
simulation design.

\begin{table}[htbp]
\centering
\caption{Simulation 2 design.}
\label{tab:sim2_parameters}
\footnotesize
\setlength{\tabcolsep}{6pt}
\renewcommand{\arraystretch}{1.12}
\begin{tabularx}{\linewidth}{@{} l X X @{}}
\toprule
\textbf{Parameter} & \textbf{Meaning} & \textbf{Value in Simulation 2} \\
\midrule
\(N\) & Number of symptoms & 9 \\
Fast-layer parameters & \(\tau_i\), \(\omega_{ij}\), and \(\gamma_i\) & Same as Simulation 1 \\
\(P_{\mathrm{base}}\) & Slow-context baseline & 0 \\
\(\kappa\) & Return-to-baseline rate & 0.20 \\
\(\sigma_P\) & Slow-field fluctuation scale & 0.04 \\
\(dt\) & Slow-field time step & 0.02 \\
\(\epsilon_t\) & Standard normal noise & \(\mathcal N(0,1)\) \\
\(\xi_t\) & Acute perturbation & 1.00 at perturbation onset; 0 otherwise \\
\(b\) & Symptom-to-context feedback & 0 \\
Pre-perturbation steps & Initial period before perturbation & 200 \\
Post-perturbation steps & Recovery period after perturbation & 750 \\
Total steps per chain & Full simulated trajectory length & 950 \\
Chains & Independent trajectories & 200 \\
\bottomrule
\end{tabularx}
\end{table}

\begin{figure}[htbp]
\centering
\includegraphics[width=0.75\textwidth]{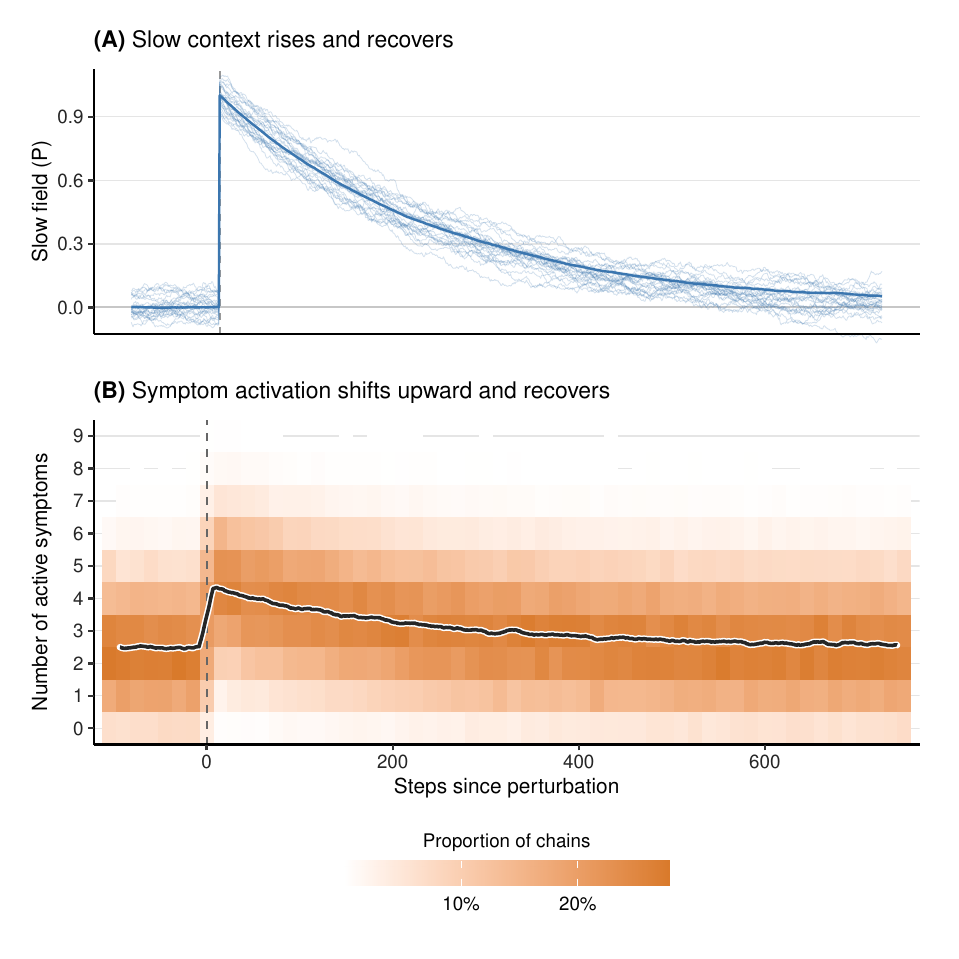}
\caption{\textbf{Recovery after a perturbation to the slow contextual field.}
(A) The slow contextual field \(P_t\) increases at perturbation onset and then
returns toward baseline. Faint lines show a sample of individual trajectories;
the bold line shows the mean across 200 chains. (B) Symptom activation shifts
upward after the perturbation and then returns toward pre-perturbation levels.
Panel B shows a time-varying distribution: at each time point, color intensity
indicates the proportion of chains with a given number of active symptoms.
Darker colors indicate that more chains had that symptom count at that time.
The bold line shows the mean number of active symptoms. Feedback from symptoms
to the slow field is turned off (\(b=0\)), isolating the slow-to-fast pathway.}
\label{fig:perturbation-recovery}
\end{figure}

\paragraph{Results.}
Figure~\ref{fig:perturbation-recovery} shows the response to a one-time perturbation of the slow contextual state. Before the perturbation, \(P_t\) fluctuated near its baseline (\(P_t\approx 0.00\)), and the symptom network averaged 2.48 active symptoms. At perturbation onset, \(P_t\) increased to approximately 1.00. Symptom activation rose over the following model steps, reaching a peak of about 4.73 active symptoms.

As \(P_t\) returned toward baseline, the symptom-count distribution shifted back toward pre-perturbation levels. By the end of the recovery window, \(P_t\) had returned close to baseline (\(P_t\approx 0.06\)), and the network averaged 2.58 active symptoms. Thus, a temporary perturbation to slow context produced a temporary increase in symptom activation, even though the symptom--symptom coupling matrix was unchanged.
A supplementary sensitivity analysis varied the overall strength of
symptom--symptom coupling while preserving the network topology. The qualitative recovery pattern was similar across weaker and reference coupling levels, although recovery was somewhat slower under the strongest coupling condition examined (Appendix~\ref{app:coupling-sensitivity}).

\subsection{Simulation 3: Feedback from symptoms to slow context}

Simulation 3 adds the reverse pathway from symptoms back to slow context. In Simulation 2, the contextual state \(P_t\) affected symptom activation, but symptoms did not affect \(P_t\) because feedback was turned off (\(b=0\)). Here, sustained symptom activation can also change the contextual state.

This creates a possible feedback loop \parencite{park2026feedback}. After a perturbation, a higher value of \(P_t\) makes symptoms easier to activate. If symptoms remain active, they can then keep \(P_t\) elevated. Recovery is therefore no longer only a matter of the contextual state returning to baseline. It also depends on whether symptom activation helps maintain the context that made symptoms likely in the first place.

Simulation 3 asks two questions. First, does symptom-to-context feedback slow
recovery after a perturbation? Second, when feedback is stronger, can the system become more dependent on its initial state? That is, if trajectories are run under the same parameters but start from contrasting states of the coupled system, do they eventually return to the same late state, or do they remain separated?

\paragraph{Design.}
Simulation 3 used the same model parameters and perturbation mechanism as
Simulation 2, but added symptom-to-context feedback. Feedback entered through
the term \(b(\bar m_t-m^\star)dt\) in Eq.~\eqref{eq:context_update}.

For Simulation 3, \(m^\star\) was set to the mean symptom activation in the
\(P=0\) condition from Simulation 1. Recent symptom activation \(\bar m_t\)
was computed using the exponentially weighted moving average described in
Appendix~\ref{app:model-details}, with smoothing rate
\(\lambda_m=0.05\).

We first examined recovery after the same perturbation across different feedback strengths. The feedback-off condition (\(b=0\)) provides the reference case in which symptoms do not feed back into the contextual process. We then extended this comparison across a broader range of feedback strengths while holding all other parameters fixed, including the fast symptom network, perturbation size, return-to-baseline rate, background fluctuation scale, and time step. The feedback values and simulation settings for the two recovery analyses are summarized in Table~\ref{tab:sim3_parameters}.

We then ran a separate initial-state check. The perturbation-recovery sweep
shows how the system responds when the same perturbation is applied across
feedback-strength conditions, but it does not show whether the coupled system
can retain a dependence on its earlier state. To examine this, we simulated
trajectories without an acute perturbation from two contrasting initial
conditions. In the low initial state, all symptoms were inactive,
\(P_0=0\), and \(\bar m_0=0\); in the high initial state, all symptoms were
active, \(P_0=1\), and \(\bar m_0=1\). All model parameters were otherwise
identical.
We then compared mean symptom activation over the final 250 steps between trajectories started from the two initial conditions. A difference near zero indicates reconvergence to the same late level, whereas a persistent difference indicates that the later state still depends on the system's initial condition.
Table~\ref{tab:sim3_parameters} summarizes the Simulation 3 settings.

\begin{table}[htbp]
\centering
\caption{Simulation 3 design.}
\label{tab:sim3_parameters}
\footnotesize
\setlength{\tabcolsep}{6pt}
\renewcommand{\arraystretch}{1.12}
\begin{tabularx}{\linewidth}{@{} l X X @{}}
\toprule
\textbf{Parameter} & \textbf{Meaning} & \textbf{Value in Simulation 3} \\
\midrule
Fast-layer parameters
  & \(\tau_i\), \(\omega_{ij}\), and \(\gamma_i\)
  & Same as Simulations 1--2 \\

Perturbation size
  & Acute perturbation to \(P_t\)
  & Same as Simulation 2 \\

\(\kappa\)
  & Return-to-baseline rate
  & 0.20 \\

\(\sigma_P\)
  & Contextual fluctuation scale
  & 0.04 \\

\(dt\)
  & Slow-process time step
  & 0.02 \\

\(\bar m_t\)
  & Smoothed recent symptom activation
  & Exponentially weighted moving average \\

\(\lambda_m\)
  & Smoothing rate
  & 0.05 \\

\(m^\star\)
  & Feedback reference level
  & \(0.277\), corresponding to 2.49 active symptoms in the \(P=0\)
    condition from Simulation 1 \\

Main recovery comparison
  & Moderate-feedback comparison
  & \(b=0\) versus \(b=0.50\); 750 post-perturbation steps;
    1,000 chains per condition \\

Feedback-strength sweep
  & Broader recovery analysis
  & \(b\in\{0,0.50,0.75,0.90,1.00,1.10,1.25,1.30,1.50\}\);
    2,500 post-perturbation steps; 300 chains per condition \\

\(b\), initial-state check
  & Feedback strengths used to assess initial-state dependence
  & \(0\) to \(1.5\) in increments of \(0.1\) \\

Initial-state check
  & Contrasting initial conditions and simulation length
  & Low: \(S_{i,0}=0\), \(P_0=0\), \(\bar m_0=0\);
    high: \(S_{i,0}=1\), \(P_0=1\), \(\bar m_0=1\);
    1,500 steps; 200 chains per initial condition and \(b\);
    final 250 steps compared \\

\bottomrule
\end{tabularx}
\end{table}

\begin{figure}[htbp]
\centering
\includegraphics[width=1.1\textwidth]{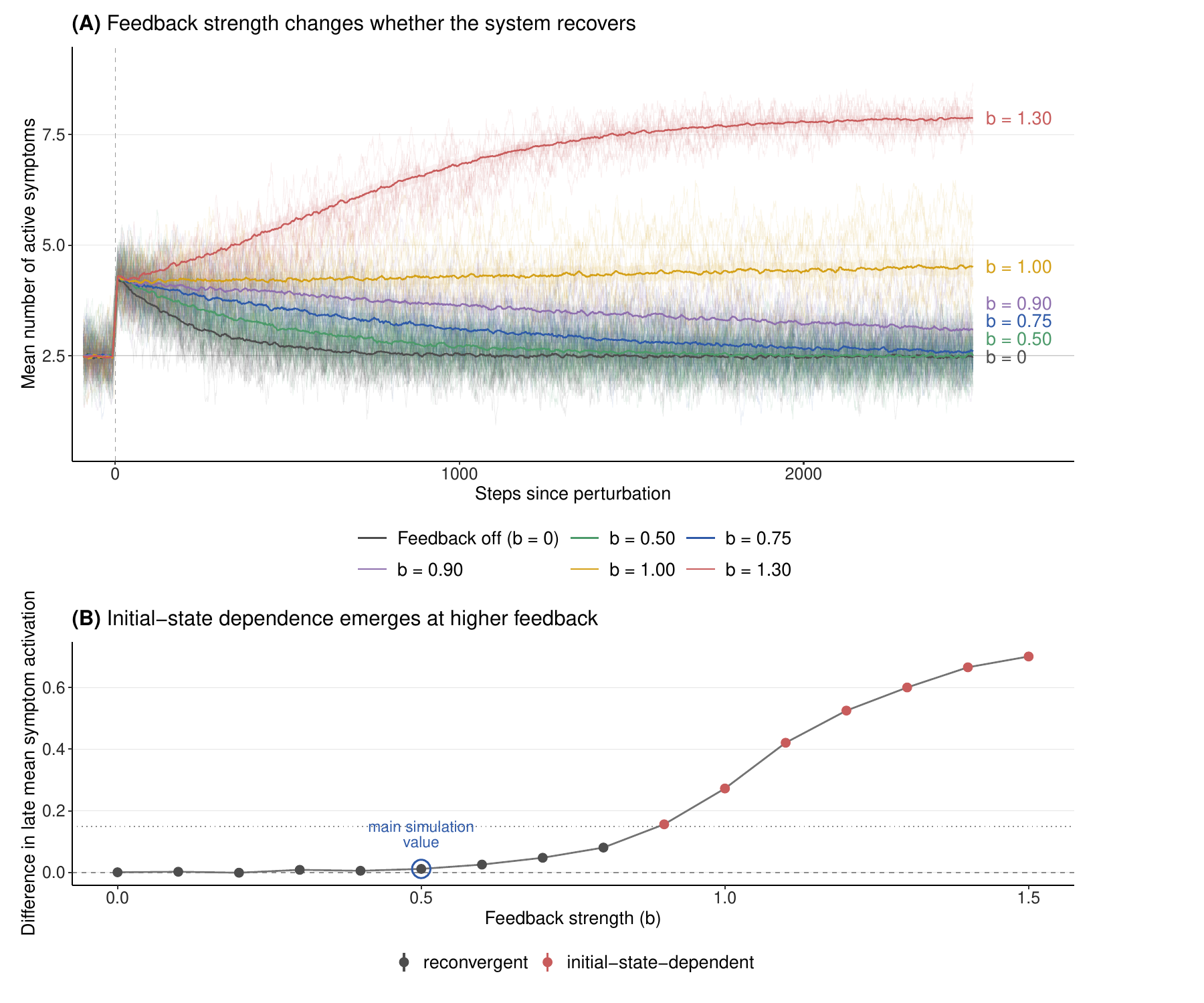}
\caption{\textbf{Symptom-to-context feedback slows recovery and can create
initial-state dependence.}
(A) Recovery after the same perturbation across feedback strengths. Six
representative feedback strengths from the nine-value sweep are shown. With
feedback off (\(b=0\)), symptom activation returns toward its pre-perturbation level. With moderate feedback (\(b=0.50\)), recovery is slower but still reconvergent. At stronger feedback values, especially around and above \(b=0.90\), symptom activation remains elevated within the simulated recovery window. At the highest value shown (\(b=1.30\)), symptom activation continues to increase after the perturbation. (B) Initial-state check. For each feedback strength, trajectories were initiated from contrasting low and high states of the coupled system, using the same parameter settings and without an acute perturbation. The low initial condition had all symptoms inactive, \(P_0=0\), and \(\bar m_0=0\); the high initial condition had all symptoms active, \(P_0=1\), and \(\bar m_0=1\). 
Values near zero mean that the two sets of trajectories reconverged to the
same late level of symptom activation. Larger values mean that late symptom
activation still depended on the initial condition. The moderate-feedback
value used in the main recovery comparison (\(b=0.50\)) remains in the
reconvergent range.}
\label{fig:recovery-feedback}
\end{figure}

\paragraph{Results.}
Figure~\ref{fig:recovery-feedback} summarizes how recovery changes as
symptom-to-context feedback increases and how stronger feedback relates to
initial-state dependence. Before the perturbation, the feedback-off and
moderate-feedback conditions were both close to baseline. With feedback off,
the mean slow field during the last 50 pre-perturbation steps was approximately
\(P_t=-0.003\), and the symptom network averaged 2.50 active symptoms. With
moderate feedback (\(b=0.50\)), the corresponding values were \(P_t=-0.002\)
and 2.49 active symptoms.

The immediate response to the perturbation was similar across conditions. In
the first 20 post-perturbation steps, peak activation was 4.34 active symptoms
with feedback off and 4.38 with moderate feedback. The difference emerged during
recovery. By the end of the recovery window, the slow field had nearly returned
to baseline when feedback was off (\(P_t=0.05\)), but remained higher with
moderate feedback (\(P_t=0.23\)). Symptom activation showed the same pattern:
the network averaged 2.57 active symptoms with feedback off, compared with 2.88
active symptoms with moderate feedback.

Panel A shows that this pattern became stronger as feedback increased. At the
moderate value used in the main comparison (\(b=0.50\)), the system still
recovered toward its pre-perturbation level. At stronger feedback values,
especially around and above \(b=0.90\), symptom activation did not return to its
pre-perturbation level over the simulated recovery period. At the highest value
shown (\(b=1.30\)), symptom activation continued to rise after the perturbation
rather than returning toward baseline.

Panel B shows a transition from reconvergence to initial-state dependence as feedback increased. Around \(b\approx0.90\), trajectories starting from the contrasting low and high initial conditions no longer reconverged to the same late level of symptom activation.
In this regime, later symptom activation depended on where the system started. The moderate-feedback value used in the main recovery comparison (\(b=0.50\)) remained below this regime, indicating that initial-state dependence is a behavior the model can exhibit under stronger feedback, not an assumption built into the main recovery result.

\subsection{Simulation 4: What estimated networks recover}

The final simulation turns from the data-generating process to the estimation
problem. The previous simulations used the slow--fast model to generate symptom trajectories. Here, we ask how data generated by the same model are represented when analyzed with an estimated symptom network.

If the slow contextual field affects symptom activation but is not included in
the estimated model, some of its influence may be represented as stronger symptom--symptom relations. Simulation 4 tests this possibility by comparing estimated networks when the slow context is held fixed, when it varies but is left out of the estimator, and when the same varying context is included in the estimation.

\paragraph{Design.}
We generated cross-sectional symptom profiles from the fast symptom model
conditional on person-level contextual values \(P_i\). Each simulated person
contributed one symptom profile and one fixed contextual value; there was no
within-person evolution of \(P_t\) in this simulation. The symptom-specific
parameters \(\tau_i\), \(\omega_{ij}\), and \(\gamma_i\) were the same as in
Simulations 1--3, so the data-generating coupling matrix \(\omega\) was fixed
and known.

We compared three estimation conditions. In the fixed-\(P\) condition, every
simulated person had the same contextual value, \(P_i=0\). In the two
varying-\(P\) conditions, each person was assigned one fixed contextual value,
\[
P_i \sim \mathrm{Uniform}(-0.6,0.6),
\]
and one symptom profile was generated conditional on that value. Contextual
variation therefore occurred between people rather than within people over
time. The distribution of \(P_i\) was centered at zero so that the
varying-\(P\) conditions had the same average contextual level as the
fixed-\(P\) condition while introducing between-person contextual
heterogeneity.

The \(P\)-omitted and \(P\)-included conditions used exactly the same simulated
people, symptom profiles, and person-level values of \(P_i\). In the
\(P\)-omitted condition, each symptom was estimated from the other symptom
variables only. In the \(P\)-included condition, \(P_i\) was additionally
included as a covariate in each nodewise logistic regression. The two
conditions therefore differed only in whether person-level context was
represented in the estimation model.

Networks were estimated using nodewise logistic regressions and symmetrized to
obtain one estimated coupling per symptom pair. We repeated the full procedure
across 30 independent replicates, with 10,000 simulated people per condition
in each replicate.

We used two primary summaries. Estimated total coupling was defined as the
signed sum of the estimated pairwise couplings,
\[
\sum_{i<j}\hat\omega_{ij}.
\]
Because all nonzero data-generating couplings were positive, the corresponding
data-generating value was \(3.05\). Using the signed sum avoids the systematic
upward contribution that arises when sampling variation around truly zero
edges is converted to positive values by taking absolute values. We separately
quantified spurious coupling among truly uncoupled symptom pairs as
\[
\sum_{\omega_{ij}=0}|\hat\omega_{ij}|.
\]
Table~\ref{tab:sim4_parameters} summarizes the design.

\begin{table}[htbp]
\centering
\caption{Simulation 4 design.}
\label{tab:sim4_parameters}
\footnotesize
\setlength{\tabcolsep}{6pt}
\renewcommand{\arraystretch}{1.12}
\begin{tabularx}{\linewidth}{@{} l X X @{}}
\toprule
\textbf{Parameter} & \textbf{Meaning} & \textbf{Value in Simulation 4} \\
\midrule
Fast-layer parameters
  & \(\tau_i\), \(\omega_{ij}\), and \(\gamma_i\)
  & Same as Simulations 1--3 \\

Data type
  & Observation structure
  & One symptom profile and one fixed \(P_i\) per simulated person \\

Fixed-\(P\) condition
  & Person-level contextual value
  & \(P_i=0\) for every simulated person \\

\(P\)-omitted condition
  & Context varies between people but is omitted from estimation
  & \(P_i\sim\mathrm{Uniform}(-0.6,0.6)\) \\

\(P\)-included condition
  & Same simulated data, with person-level context included in estimation
  & \(P_i\sim\mathrm{Uniform}(-0.6,0.6)\) \\

Network estimator
  & Estimation method
  & Nodewise logistic regressions, symmetrized across symptom pairs \\

Replicates
  & Independent simulation replicates
  & 30 \\

Profiles per condition
  & Simulated people per condition per replicate
  & 10,000 \\

Estimated total coupling
  & Signed sum of estimated pairwise couplings
  & \(\sum_{i<j}\hat\omega_{ij}\) \\

Spurious coupling
  & Absolute estimated coupling among truly uncoupled pairs
  & \(\sum_{\omega_{ij}=0}|\hat\omega_{ij}|\) \\
\bottomrule
\end{tabularx}
\end{table}

\begin{figure}[htbp]
\centering
\includegraphics[width=0.78\textwidth]{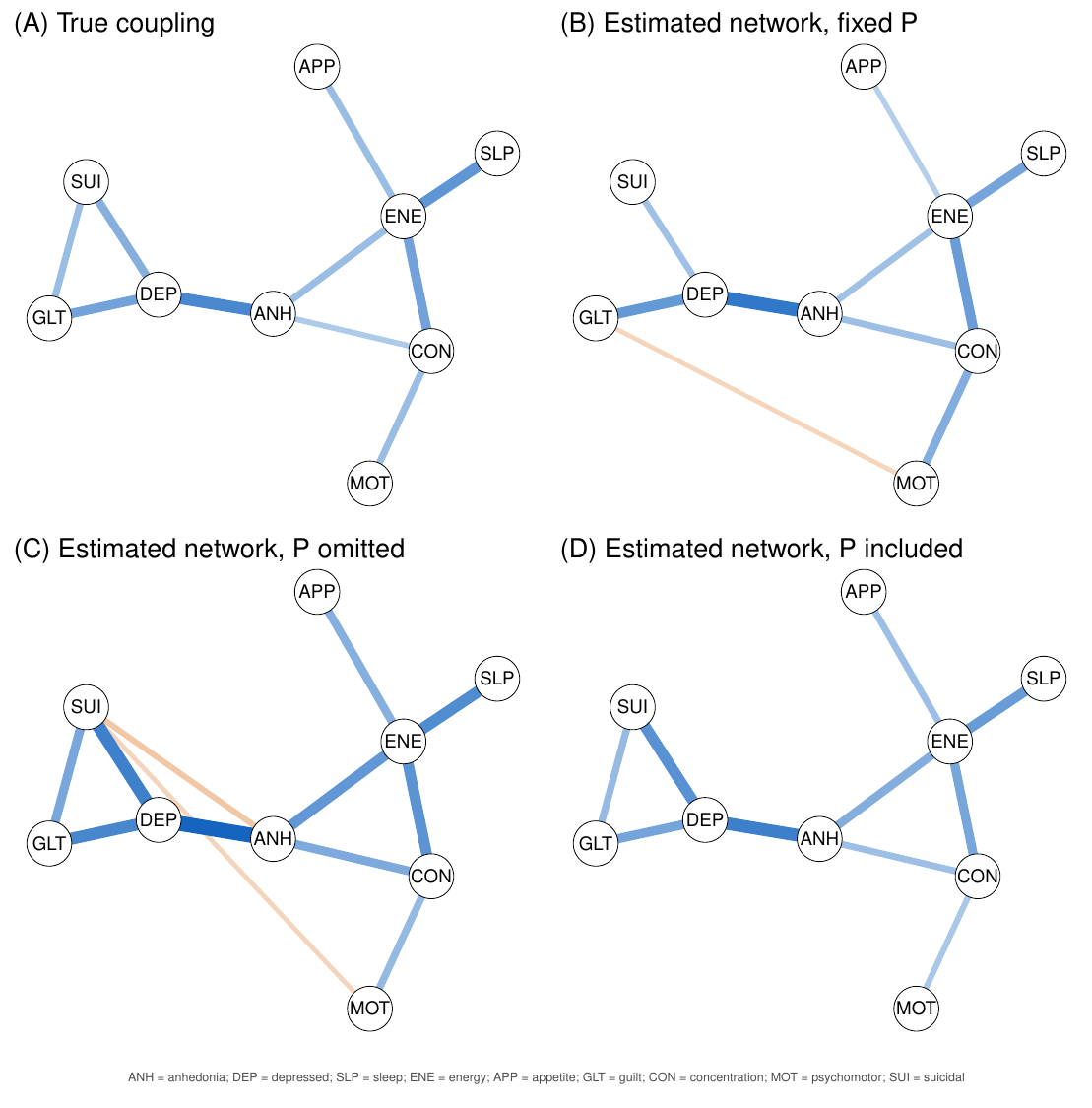}
\vspace{0.75em}
\includegraphics[width=1\textwidth]{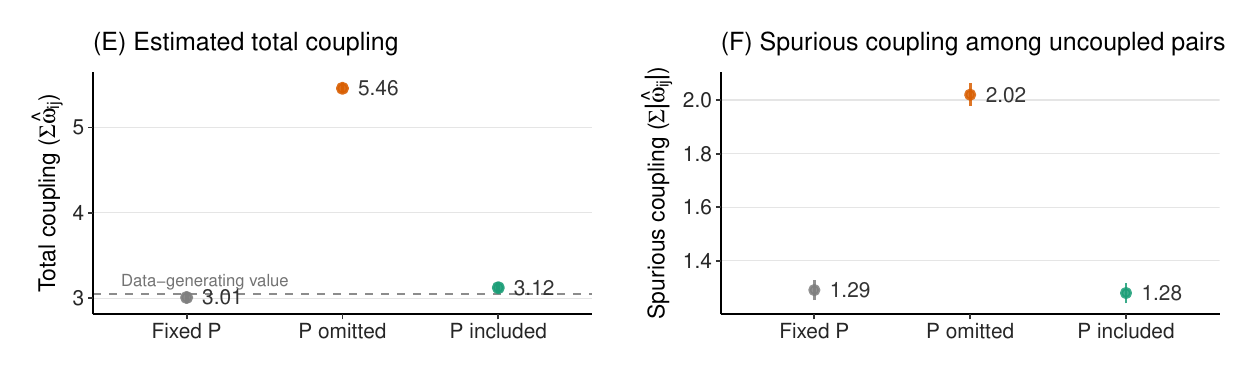}
\caption{\textbf{Omitting slow context increases estimated symptom coupling.}
(A) Data-generating symptom--symptom coupling matrix.
(B) Network estimated when every simulated person had the same contextual
value, \(P_i=0\).
(C) Network estimated from pooled symptom profiles when person-level contextual
values varied across people but were omitted from the estimation model.
(D) Network estimated from the same pooled data when \(P_i\) was included in
the estimation model. In Panels B--D, orange edges indicate estimated
couplings exceeding the plotting threshold of \(0.15\) for symptom pairs whose
data-generating coupling was zero. Panels B--D show one representative
\(n=10{,}000\) realization.
(E) Estimated total coupling, defined as the signed sum of estimated pairwise
couplings; the dashed line marks the data-generating value of \(3.05\).
(F) Spurious absolute coupling summed across symptom pairs with zero
data-generating coupling. Points and error bars in Panels E--F show means and
standard errors across 30 independent replicates.}
\label{fig:network-estimation}
\end{figure}

\paragraph{Results.}
Figure~\ref{fig:network-estimation} compares the data-generating coupling
matrix with networks estimated under the three conditions. Panel A shows the
true symptom--symptom couplings. Panels B--D show representative network
estimates when \(P_i\) was fixed, varied across people but omitted from
estimation, or varied and was included in estimation, respectively.

The visual difference is clearest when \(P_i\) varied across people but was
omitted. In Panel C, several symptom pairs with zero data-generating coupling
received nonzero estimated couplings. Orange edges in Panels B--D indicate
estimated couplings exceeding the plotting threshold of \(0.15\) for symptom
pairs whose data-generating coupling was zero.

Panels E and F quantify this pattern across 30 independent replicates. Panel E
shows estimated total coupling, defined as the signed sum of estimated pairwise
couplings. The data-generating value was \(3.05\). Estimated total coupling was
\(3.01\) (\(SE=0.07\)) in the fixed-\(P\) condition and \(3.12\)
(\(SE=0.06\)) when \(P_i\) was included, but increased to \(5.46\)
(\(SE=0.06\)) when \(P_i\) varied across people and was omitted from the
estimation model.

Panel F isolates estimated coupling among symptom pairs that were uncoupled in
the data-generating model. The summed absolute coupling across these true-zero
pairs was \(1.29\) (\(SE=0.04\)) in the fixed-\(P\) condition,
\(1.28\) (\(SE=0.04\)) when \(P_i\) was included, and \(2.02\)
(\(SE=0.04\)) when \(P_i\) was omitted. Within the paired varying-\(P\)
replicates, omitting \(P_i\) increased estimated total coupling by \(2.33\)
(\(SE=0.03\)) and spurious coupling on true-zero pairs by \(0.74\)
(\(SE=0.03\)); both differences were positive in all 30 replicates.

Thus, between-person contextual variation produced additional estimated
symptom--symptom coupling when it was omitted from the estimation model.
Including \(P_i\) brought estimated total coupling close to the
data-generating value and reduced spurious coupling on truly uncoupled symptom
pairs to approximately the level observed when context did not vary across
people.

\section{Discussion}

The central implication of this paper is interpretive. Estimated symptom
networks are often read as if most of the relevant action lies among symptoms.
The model developed here suggests a wider reading. The symptom layer is one
part of a coupled process. Symptom--symptom coupling still matters, but what it
produces also depends on the slower contextual process in which the symptom
system is embedded.

This matters for how we read activation, recovery, and persistence. A higher
number of active symptoms may reflect stronger symptom--symptom coupling, but
it may also mean that several symptoms have become easier to activate at once.
Recovery is similar. It is not only about whether symptoms stop reinforcing one
another. It also depends on whether the conditions that made symptoms easier to
activate begin to change. If sustained symptoms help keep those conditions in
place, the system can start to resemble a vicious cycle. Symptoms are shaped by
context, but they can also help shape the context in which later symptoms
unfold.

This is the reason for modeling feedback explicitly. Without feedback, slow
context can still affect symptom activation, but recovery is largely a question
of whether the contextual state returns toward baseline. With feedback,
symptoms can become part of the process that keeps later activation likely.
Persistence therefore does not have to be located entirely inside the symptom
network. It can also arise from the loop between symptom activation and the
slower conditions that support later activation \parencite{park2024cyclic}.

The same logic carries over to estimation. If a slower contextual process
affects several symptoms but is not represented in the model, some of that
shared input can appear as additional symptom--symptom coupling. This does not
mean that estimated edges should be dismissed. Rather, estimated associations
may combine direct symptom coupling with variation induced by shared contextual
input.

The implication arises first at the level of individual dynamics. The same
symptom system can show different levels of activation, recovery, and
persistence as its slower contextual conditions change, even when
symptom--symptom coupling remains fixed. Interpreting an individual's symptom
dynamics therefore requires attention not only to relations among symptoms, but
also to slower processes that may shift their activation tendencies over time.

A related issue arises in between-group comparisons when groups differ
systematically in these contextual conditions. This is particularly relevant
for observationally defined groups, or more generally when the design does not
control the contextual processes that differ between the groups being compared.
In such cases, a more strongly connected estimated network may reflect stronger
direct symptom coupling, but it may also partly reflect differences in
contextual input or in how observations are distributed across contextual
conditions. Under successful randomization or control of the relevant
contextual differences, this particular source of between-group distortion
should be reduced. Network comparisons are therefore most informative when the
design makes clear which contextual processes are controlled and when
activation tendencies and relevant contextual variables are considered
alongside estimated edges.

At the same time, the present model is only a first step toward that kind of
interpretation. Its strength is that it makes the slow--fast mechanism visible
under controlled conditions, but this also means that many parts of real
psychological systems are simplified. In particular, the division into one
fast symptom layer and one slow contextual process should not be read as a
literal binary classification of psychological variables. Symptoms themselves
may differ in their characteristic timescales, and social, biological, and
psychological processes may operate across several overlapping timescales.
What matters in the present model is the relative separation between a faster
process and another process that changes sufficiently slowly to shape its
dynamics.

The contextual process is also one-dimensional, symptom states are binary, and
the parameter values are illustrative rather than fitted to empirical data.
Real psychological systems are likely to involve multiple slower processes,
symptom-specific rates of change, symptom-specific context loadings,
heterogeneous positive and negative couplings, and measurement processes that
differ across people and settings. The simulations should therefore be read as
theory-building examples rather than calibrated clinical predictions.

The perturbation used in the recovery simulations is also simplified. It is a
probe of recovery, not a full model of stressful life events. A richer event
model could distinguish gradual drift in slow context from discrete events that
move the system suddenly, or could represent stress generation as a
symptom-linked event process
\parencite{hammen2006stress, dohrenwend2006inventorying, cohen2019ten}.

The \(P\)-included estimator illustrates one simple methodological implication
of the model: when a known source of contextual variation is represented in the
estimation model, recovery of the underlying symptom coupling improves. In real
data, however, contextual variables will often be incomplete, noisy, or only
partially observed, so ordinary covariate adjustment will not generally be
enough to identify direct symptom--symptom effects. A broader direction is to
model contextual or intervention variables jointly with the system itself.
Related approaches such as Joint Causal Inference formalize this idea by using
variation across observational or interventional contexts to help identify
causal structure \parencite{mooij2020joint}.

These limitations mark the next step. Future work can allow multiple
interacting timescales, symptom-specific update rates, empirically measured
contextual variables, multiple slower processes, symptom-specific context
loadings, heterogeneous coupling structures, and observation models matched to
ecological momentary assessment (EMA) or clinical panel data. An important
methodological extension is to study how slow--fast processes can be identified
from repeated within-person data, where symptoms and contextual processes are
both observed over time rather than reduced to a single cross-sectional
profile. Such extensions would make it possible to ask how much of an observed
pattern is attributable to direct symptom coupling, how much to slow context,
and how much to feedback between the two.

More broadly, the slow--fast framework changes the modeling question. The
issue is not only whether symptoms influence one another, or whether context
predicts symptoms. It is how processes with different speeds become tied
together. Slow context can set the conditions of the system, change over time,
recover after perturbation, or be changed by sustained symptom activation.
These cases can produce similar symptom patterns, but they describe different
systems. Making those alternatives explicit is the value of the model.

This paper developed the framework through the case of psychopathology symptom
networks, but the problem is broader than symptom networks. Much of psychology
deals with systems in which some things change from moment to moment, while
other conditions accumulate slowly and set the terms for what can happen next.

The central idea is that these parts should not be studied only in isolation.
A fast state may pass quickly, but repeated or sustained states can change the
slower conditions that make similar states more likely later. This is the basic
slow--fast problem. It is not just that context affects psychological states,
or that states interact with one another. It is that fast and slow parts of the
system can keep changing what each other can do.

We hope this framework helps researchers study those links more directly. In
the symptom-network case, it helps clarify how symptom activation, recovery,
feedback, and estimated connectivity can belong to the same process. More
generally, it offers a way to think about psychological systems in which
different parts move at different speeds, influence one another across
timescales, and sometimes make change or recovery easier or harder.

\section{Open Practices and Data Availability}

All simulations are based on synthetic data generated from the model described in the manuscript. Simulation code, parameter settings, and figure-generation
scripts are available at \url{https://github.com/KyuriP/slow-fast-coupled-dynamics-model}.

\printbibliography

\appendix

\section{Technical details of the slow--fast model}\label{app:model-details}

This appendix provides technical details that complement the streamlined model
description in the main text. The main text focuses on the conceptual structure
of the slow--fast system and the key equations needed to interpret the
simulations. Here we collect implementation details for the fast-layer update,
the smoothed symptom signal, the slow-field update, and the acute perturbation
used in the recovery simulations.

\subsection{Fast-layer update dynamics}

At each slow time step \(t\), the slow contextual field \(P_t\) is held fixed
while the fast symptom layer is updated. The fast layer consists of \(N\) binary
symptom states,
\[
S_{i,t}\in\{0,1\},
\]
where \(S_{i,t}=0\) denotes an inactive symptom and \(S_{i,t}=1\) denotes an
active symptom.

Given the current states of the other symptoms and the current slow field, the
local input to symptom \(i\) is
\[
\eta_{i,t}
=
\tau_i+\sum_{j\neq i}\omega_{ij}S_{j,t}+\gamma_iP_t .
\]
The updated state of symptom \(i\) is then drawn from
\[
S_i^{\mathrm{new}}
\sim
\mathrm{Bernoulli}\!\left(\operatorname{logit}^{-1}(\eta_{i,t})\right).
\]
Each symptom update is stochastic and conditional on the current state of
the rest of the symptom network.

In the simulations, one fast sweep consists of updating all \(N\) symptoms once in random order. The slow field \(P_t\) is held fixed during this sweep. After a symptom is updated, its new value is used in subsequent symptom updates within the same sweep.
This gives an asynchronous local-update process rather than a synchronous update in which all symptoms are changed at the same time. The asynchronous scheme avoids imposing artificial simultaneity and more closely approximates a continuous-time process in which state changes occur sequentially, although the present implementation itself remains discrete-time.

When several fast sweeps are used within one slow time step, the symptom layer
has more opportunity to adjust to the current value of \(P_t\) before the slow
field changes again. This implements the intended timescale separation: symptom activation can fluctuate rapidly, while the contextual field changes more gradually.

\subsection{Mean and smoothed symptom activation}

The instantaneous mean symptom activation is
\[
m_t=\frac{1}{N}\sum_{i=1}^{N}S_{i,t}.
\]
This quantity is the proportion of symptoms that are active at time \(t\). The slow contextual process does not respond directly to every instantaneous value of \(m_t\).
Instead, symptom-to-context feedback uses a smoothed symptom signal,
\(\bar m_t\), which summarizes recent symptom activation.

We compute this smoothed signal using an exponentially weighted moving average:
\[
\bar m_{t+dt}
=
(1-\lambda_m)\bar m_t+\lambda_m m_t,
\qquad 0<\lambda_m<1.
\]
The smoothing parameter \(\lambda_m\) controls how quickly the smoothed symptom
signal responds to the current symptom state. Larger values make \(\bar m_t\)
track momentary symptom activation more closely, whereas smaller values make
\(\bar m_t\) depend more strongly on recent history. This is what gives the
feedback pathway its slow timescale: the slow contextual process responds to
sustained symptom activation, not to every momentary fluctuation in \(m_t\).
In Simulation 3, we set \(\lambda_m=0.05\).

Equivalently,
\[
\bar m_{t+dt}
=
\bar m_t+\lambda_m(m_t-\bar m_t).
\]
Thus, \(\bar m_t\) moves only partway toward the current value of \(m_t\) at
each slow time step.

Iterating the recursion shows that past symptom activation receives
geometrically decaying weights:
\[
\bar m_t
=
\lambda_m
\sum_{k=0}^{t-1}
(1-\lambda_m)^k m_{t-1-k}
+
(1-\lambda_m)^t\bar m_0 .
\]
Values of \(m_t\) from the recent past receive the largest weight, while older
values receive progressively smaller weights. Smaller values of \(\lambda_m\)
therefore imply longer memory and stronger separation between fast symptom
fluctuation and slow contextual change.

This smoothing is important for the interpretation of symptom-to-context
feedback. A single brief symptom spike should not immediately change a slow
contextual condition. Sustained symptom activation, however, may affect work, relationships, sleep, stress exposure, or recovery. The smoothed signal
\(\bar m_t\) is meant to capture this distinction.

\subsection{Slow-field update}

The slow contextual field is updated according to
\[
P_{t+dt}
=
P_t
+
\left[
-\kappa(P_t-P_{\mathrm{base}})
+
b(\bar m_t-m^\star)
\right]dt
+
\sigma_P\sqrt{dt}\,\epsilon_t
+
\xi_t,
\qquad
\epsilon_t\sim\mathcal N(0,1).
\]
The first term, \(P_t\), is the current level of the slow contextual field. The
term
\[
-\kappa(P_t-P_{\mathrm{base}})dt
\]
pulls the slow field back toward its baseline \(P_{\mathrm{base}}\). The
parameter \(\kappa\) controls how quickly this return occurs.

The term
\[
\sigma_P\sqrt{dt}\,\epsilon_t
\]
represents small background fluctuations in the slow field. Here,
\(\epsilon_t\) is standard normal noise and \(\sigma_P\) controls the scale of
the fluctuation. The factor \(\sqrt{dt}\) is used because this term represents a
diffusion-like fluctuation over a time step of length \(dt\).

The feedback term
\[
b(\bar m_t-m^\star)dt
\]
allows sustained symptom activation to affect the slow field. The parameter
\(m^\star\) is the reference level of recent symptom activation. When
\(\bar m_t>m^\star\), feedback pushes the slow field upward. When
\(\bar m_t<m^\star\), feedback pushes the slow field downward. The parameter
\(b\) controls the strength of this feedback.

Finally, \(\xi_t\) represents occasional larger perturbations to the slow field.
Unlike the diffusion term, \(\xi_t\) is not multiplied by \(dt\). It represents a
discrete jump in the slow field when an acute event occurs.

\subsection{General stochastic jump process for acute perturbations}

In the recovery simulations reported in the main text, the jump term
\(\xi_t\) is implemented as a single deterministic perturbation:
\(\xi_t=1\) at perturbation onset and \(\xi_t=0\) otherwise. More generally,
the same jump term can represent stochastic acute events with their own timing
and magnitude. We describe one such extension here.

Rare acute events are represented as additive jump increments \(\xi_t\) in the
slow-field update. Conceptually, the jump term is meant to capture discrete,
identifiable events with their own timing and effect size, such as conflict,
loss, financial setback, health problems, relationship change, birth, or death
\parencite{cohen2019ten, dohrenwend2006inventorying}. This differs from the
Gaussian fluctuation term, which represents the accumulation of many smaller,
unidentified background influences whose individual timing and magnitude are not
distinguished.

In this generalized formulation, acute events occur according to a rare-event process. The total event rate is
\[
\lambda_{\mathrm{tot}}(t)
=
\lambda_0
+
\lambda_1[\bar m_t-m_{\mathrm{crit}}]_+,
\qquad
[x]_+\equiv \max(0,x),\quad \lambda_1\ge 0.
\]
Here, \(\lambda_0\) is the baseline event rate. The second term allows event
risk to increase when recent symptom activation exceeds a critical level
\(m_{\mathrm{crit}}\). This captures the idea that symptoms may not only respond
to stressful conditions, but may also increase later exposure to stressors,
consistent with stress-generation accounts
\parencite{hammen2006stress, liu2010stress, rnic2023vicious}.

To simulate this process on a time grid of size \(dt\), we convert the
continuous-time event rate into the probability that at least one event occurs
during the next time step:
\[
p_{\mathrm{tot}}(t)
=
1-\exp[-\lambda_{\mathrm{tot}}(t)dt],
\qquad
I_t\sim\mathrm{Bernoulli}(p_{\mathrm{tot}}(t)).
\]
We interpret \(I_t=1\) as an event occurring during the interval
\([t,t+dt)\), and \(I_t=0\) as no event. For sufficiently small \(dt\), this
approximation allows at most one event per step while keeping the probability
of multiple events within a single step negligible.

This construction also clarifies why \(\xi_t\) is added directly to the slow
field rather than multiplied by \(dt\). The probability of an event occurring
during a step scales with \(dt\), through \(p_{\mathrm{tot}}(t)\), but the jump
size itself does not. Thus, as the time grid is refined, diffusion increments
become smaller per step, whereas a realized acute perturbation remains a finite
displacement of the slow field.

Conditional on \(I_t=1\), we label the event as baseline versus
symptom-linked in proportion to their contributions to the total rate:
\[
\Pr(\mathrm{symptom\mbox{-}linked}\mid I_t=1)
=
\frac{\lambda_1[\bar m_t-m_{\mathrm{crit}}]_+}
{\lambda_{\mathrm{tot}}(t)},
\]
and
\[
\Pr(\mathrm{baseline}\mid I_t=1)
=
\frac{\lambda_0}{\lambda_{\mathrm{tot}}(t)}.
\]
We then draw an event-type-specific magnitude and set the jump increment for
that step:
\[
\xi_t=
\begin{cases}
A_{\mathrm{base},t}, & \text{baseline event},\\
A_{\mathrm{symp},t}, & \text{symptom-linked event},\\
0, & I_t=0,
\end{cases}
\]
with
\[
A_{\mathrm{base},t}\sim
\mathcal N(\mu_{\mathrm{base}},\sigma_{\mathrm{base}}^2),
\qquad
A_{\mathrm{symp},t}\sim
\mathcal N(\mu_{\mathrm{symp}},\sigma_{\mathrm{symp}}^2).
\]

Within this generalized formulation, setting \(\xi_t=0\) removes acute perturbations. Setting \(\lambda_1=0\) removes symptom-linked event generation, so that stochastic jumps can occur only through the baseline event rate \(\lambda_0\).

This stochastic event process provides a more general implementation of the
jump term but is not used in Simulations 1--4. Those simulations use either
\(\xi_t=0\) or the single deterministic perturbation described above. The
stochastic formulation shows how acute-event timing and symptom-linked event
generation could be incorporated within the same slow--fast framework.

\subsection{Parameter values used in the main simulations}
\label{app:main-parameters}

The symptom-specific parameters used in the main simulations were specified
for theoretical illustration rather than estimated from empirical data.
The same values of \(\tau_i\), \(\gamma_i\), and \(\omega_{ij}\) were used
throughout Simulations 1--4 unless otherwise stated.

\begin{table}[htbp]
\centering
\caption{Symptom-specific baseline and slow-context loading parameters.}
\label{tab:main_symptom_parameters}
\footnotesize
\begin{tabular}{lcc}
\toprule
\textbf{Symptom} & \(\boldsymbol{\tau_i}\) & \(\boldsymbol{\gamma_i}\) \\
\midrule
Anhedonia     & -1.10 & 0.90 \\
Depressed mood & -1.00 & 1.00 \\
Sleep          & -0.70 & 0.70 \\
Energy         & -0.60 & 0.80 \\
Appetite       & -1.20 & 0.70 \\
Guilt          & -1.50 & 1.00 \\
Concentration  & -1.00 & 0.80 \\
Psychomotor    & -1.70 & 0.70 \\
Suicidal ideation & -2.70 & 0.60 \\
\bottomrule
\end{tabular}
\end{table}

\begin{table}[htbp]
\centering
\caption{Nonzero symptom--symptom couplings used in the main simulations.}
\label{tab:main_couplings}
\footnotesize
\begin{tabular}{lc}
\toprule
\textbf{Symptom pair} & \(\boldsymbol{\omega_{ij}}\) \\
\midrule
Anhedonia -- depressed mood       & 0.45 \\
Depressed mood -- guilt           & 0.35 \\
Depressed mood -- suicidal ideation & 0.30 \\
Sleep -- energy                   & 0.40 \\
Energy -- concentration           & 0.35 \\
Concentration -- psychomotor      & 0.25 \\
Appetite -- energy                & 0.25 \\
Guilt -- suicidal ideation        & 0.25 \\
Anhedonia -- energy               & 0.25 \\
Anhedonia -- concentration        & 0.20 \\
\bottomrule
\end{tabular}

\begin{tablenotes}[flushleft]
\footnotesize
\item \textit{Note.} All symptom pairs not listed in the table have
\(\omega_{ij}=0\).
\end{tablenotes}
\end{table}

\subsection{Simulation 1 post-burn-in stability check}
\label{app:sim1-stability}

\begin{figure}[htbp]
\centering
\includegraphics[width=\linewidth]{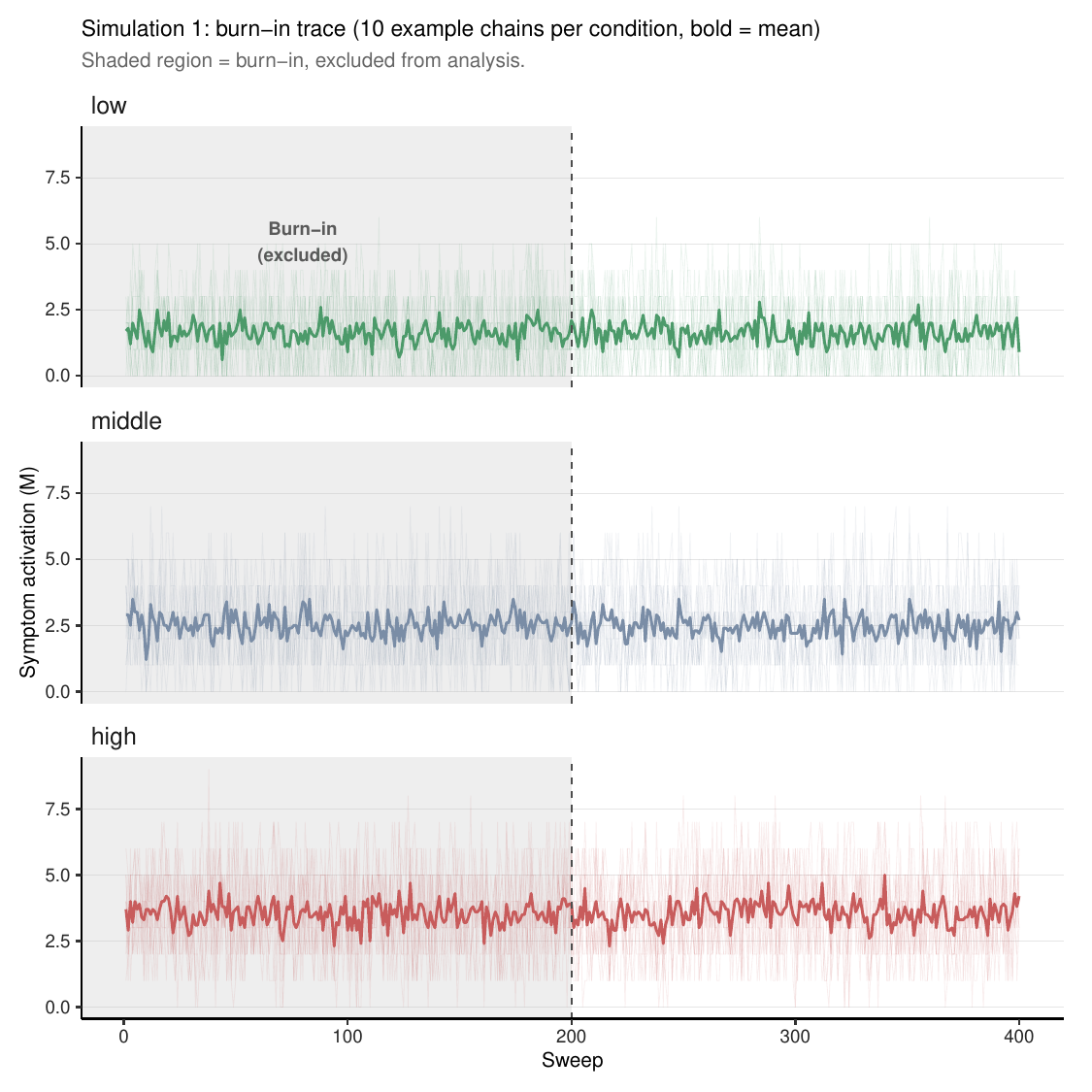}
\caption{\textbf{Burn-in traces for Simulation 1.}
Ten example chains are shown for each contextual condition. The dashed vertical
line marks the end of the 200-sweep burn-in period. The trajectories fluctuate
around condition-specific levels without an evident systematic transient
approaching or following the end of burn-in.}
\label{fig:sim1-burnin}
\end{figure}

We checked whether the 200-sweep burn-in used in Simulation 1 left systematic
drift in symptom activation. First, we inspected full trajectories from 10
example chains in each contextual condition. As shown in
Figure~\ref{fig:sim1-burnin}, the trajectories showed no visible transient or
systematic trend approaching the end of the burn-in period.

We also compared mean symptom activation during the first and second halves of
the retained 200-sweep window. The differences were \(0.0286\), \(0.0160\),
and \(0.0019\) active symptoms in the low-, reference-, and high-context
conditions, respectively. For comparison, the difference in overall mean
activation between the low- and high-context conditions was approximately
\(1.92\) active symptoms. Residual drift after burn-in was therefore small
relative to the contextual difference examined in Simulation 1.

\begin{table}[htbp]
\centering
\caption{Post-burn-in stability check for Simulation 1.}
\label{tab:sim1_stability}
\footnotesize
\begin{tabular}{lccc}
\toprule
\textbf{Condition}
& \textbf{First 100 sweeps}
& \textbf{Second 100 sweeps}
& \textbf{Difference} \\
\midrule
Low context (\(P=-0.6\)) & 1.6256 & 1.6542 & 0.0286 \\
Reference (\(P=0\))      & 2.4818 & 2.4978 & 0.0160 \\
High context (\(P=0.6\)) & 3.5559 & 3.5577 & 0.0019 \\
\bottomrule
\end{tabular}
\end{table}

\subsection{Sensitivity of recovery to symptom-coupling strength}
\label{app:coupling-sensitivity}

Simulation 2 holds the symptom--symptom coupling matrix fixed in order to
isolate recovery following a perturbation to the contextual process. As a
sensitivity check, we repeated the same perturbation-and-recovery simulation
while varying the overall strength of symptom--symptom coupling.

We preserved the topology and relative edge weights of the original coupling
matrix and rescaled all nonzero couplings by a common factor,
\[
\omega^{(c)} = c\omega.
\]
The full sensitivity grid used
\[
c\in\{0.3,0.5,1.0,1.5,2.0,2.5,3.0\},
\]
with \(c=1\) corresponding to the coupling matrix used in the main simulations.
All other parameters, including \(\tau_i\), \(\gamma_i\), the slow-context
dynamics, and the perturbation magnitude, were held fixed. Feedback from
symptoms to context remained off (\(b=0\)), as in Simulation 2. We simulated
1,000 independent trajectories for each coupling condition.

Because changing coupling strength also changes the baseline level of symptom
activation, recovery was evaluated relative to each condition's own
pre-perturbation level. We defined normalized excess activation as
\[
R(t)
=
\frac{\bar M(t)-\bar M_{\mathrm{pre}}}
     {\bar M_{\mathrm{peak}}-\bar M_{\mathrm{pre}}},
\]
where \(\bar M_{\mathrm{pre}}\) is the mean symptom activation during the final
50 pre-perturbation steps and \(\bar M_{\mathrm{peak}}\) is the maximum mean
activation during the first 20 post-perturbation steps. Thus, \(R(t)=1\)
corresponds to the initial perturbation response and \(R(t)=0\) to return to
the condition-specific pre-perturbation level.

\begin{figure}[htbp]
\centering
\includegraphics[width=\linewidth]{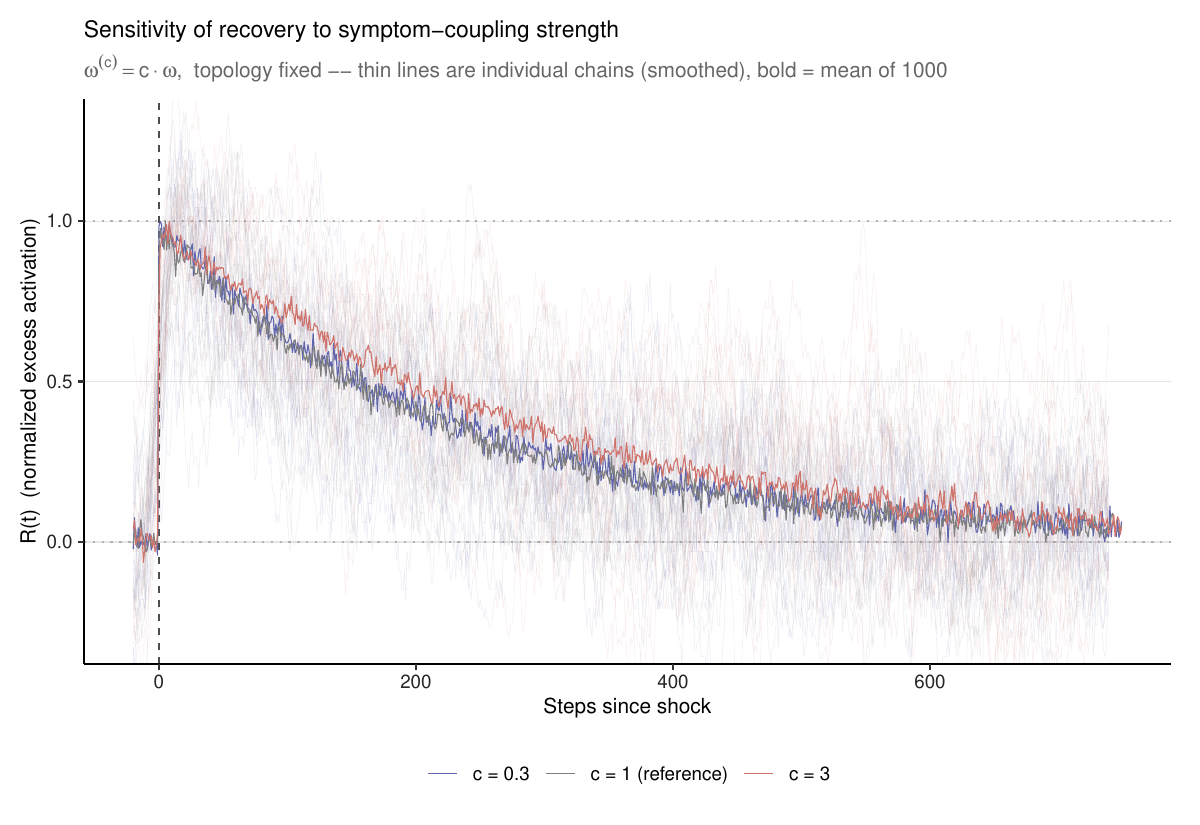}
\caption{\textbf{Sensitivity of recovery to symptom--symptom coupling strength.}
The data-generating coupling matrix was rescaled as
\(\omega^{(c)}=c\omega\) while its topology and relative edge weights were
held fixed. Curves show recovery relative to each coupling condition's own
pre-perturbation activation level. Thin lines show 15 smoothed example
trajectories per condition; bold lines show the mean across 1,000 independent
trajectories. The dashed vertical line marks perturbation onset.}
\label{fig:coupling-sensitivity}
\end{figure}

Figure~\ref{fig:coupling-sensitivity} shows three representative coupling
levels from the broader grid. Recovery was very similar under weaker coupling
(\(c=0.3\)) and the reference coupling (\(c=1\)), whereas the strongest
condition shown (\(c=3\)) retained somewhat more excess activation during
recovery. The trajectories nevertheless moved back toward their own
pre-perturbation levels over the simulated recovery window. This check
indicates that the qualitative recovery pattern in Simulation 2 is not
specific to the exact coupling magnitude used in the reference model, while
also showing that sufficiently strong symptom coupling can modestly prolong
the response to the same contextual perturbation.

\section{Supplementary NCT check under threshold differences only}\label{app:nct}

This supplementary check is not intended to match the \(N=9\) PHQ-style
simulation design used in the main text. It uses \(N=12\) symptoms to examine a
more general methodological question: whether standard network-comparison
procedures can flag network differences when two groups differ only in
thresholds and share the same data-generating coupling matrix. The check should
therefore be read as a standalone robustness analysis about network comparison
under threshold differences, not as a direct extension of Simulations 1--4.

To examine whether standard network comparison procedures absorb threshold
differences into threshold parameters rather than into edge estimates, we
conducted a supplementary Monte Carlo analysis. In each condition, we simulated
two groups from Ising models with the same coupling matrix but different
threshold vectors, estimated the networks separately with \texttt{IsingFit},
and compared them using the weighted Network Comparison Test (NCT). Thus, the
two groups were generated from the same underlying symptom network and differed
only in their baseline tendency for symptoms to become active.

The true network topology was varied across three cases: dense, modular, and
ring. These are not participant groups, but three different coupling structures
used to generate the data. Within each condition, both groups shared the same
coupling matrix and differed only in thresholds. Table~\ref{tab:nct_design}
summarizes the supplementary NCT design.

\begin{table}[htbp]
\centering
\begin{threeparttable}
\caption{Design of the supplementary NCT simulation.}
\label{tab:nct_design}
\footnotesize
\setlength{\tabcolsep}{6pt}
\renewcommand{\arraystretch}{1.15}
\begin{tabularx}{\linewidth}{@{} l X @{}}
\toprule
\textbf{Category} & \textbf{Setting} \\
\midrule
Symptoms & \(N=12\) binary symptoms. \\
Groups & Two groups per replicate. The groups had identical coupling and differed only in thresholds. \\
Dense topology & Fully connected 12-node network: all off-diagonal edge weights set to 0.25, diagonal entries set to 0. \\
Modular topology & Two 6-node modules: all within-module off-diagonal edge weights set to 0.25, all between-module edge weights set to 0.05, diagonal entries set to 0. \\
Ring topology & 12-node ring: each node connected only to its two immediate neighbors with edge weight 0.25; all other off-diagonal entries set to 0. \\
Low-threshold group & All 12 thresholds set to \(-2.0\). \\
High-threshold group & All 12 thresholds set to \(-2.0+\Delta\), where \(\Delta \in \{0.4, 0.8, 1.2\}\). Thus the second group used thresholds \(-1.6\), \(-1.2\), or \(-0.8\). \\
Meaning of threshold difference & Less negative thresholds imply a higher baseline probability that symptoms are active. \\
Sample size per group & \(n \in \{200, 500, 1000\}\). \\
Monte Carlo replicates & 100 replicates per condition. \\
Data generation & \texttt{IsingSampler}, binary responses \(\{0,1\}\), Metropolis--Hastings sampler. \\
Network estimation & \texttt{IsingFit} with \texttt{family = "binomial"}. \\
Network comparison & Weighted NCT. \\
Permutations per NCT run & 250. \\
\bottomrule
\end{tabularx}
\end{threeparttable}
\end{table}

We summarized the results in three ways. First, we recorded how often NCT
rejected the structure invariance test, that is, how often it concluded that
the estimated edge-weight configuration differed between the two groups. Second,
we recorded how often NCT rejected the global strength test. Third, because we
requested edge-wise tests, we computed the average proportion of individual
edges falsely flagged as significant at \(p<.05\). Under the null hypothesis of
identical coupling, all three quantities would ideally remain close to the
nominal 5\% level.

\begin{figure}[htbp]
  \centering
  \includegraphics[width=\linewidth]{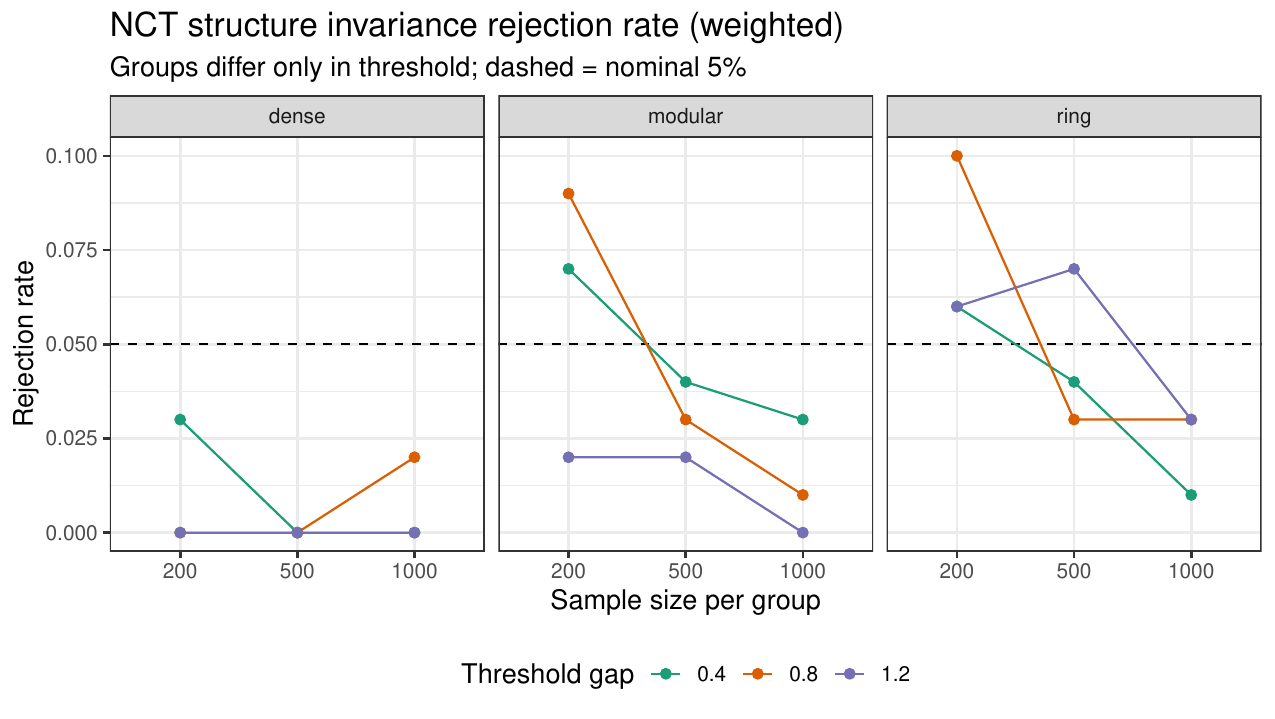}
  \vspace{0.8em}
  \includegraphics[width=\linewidth]{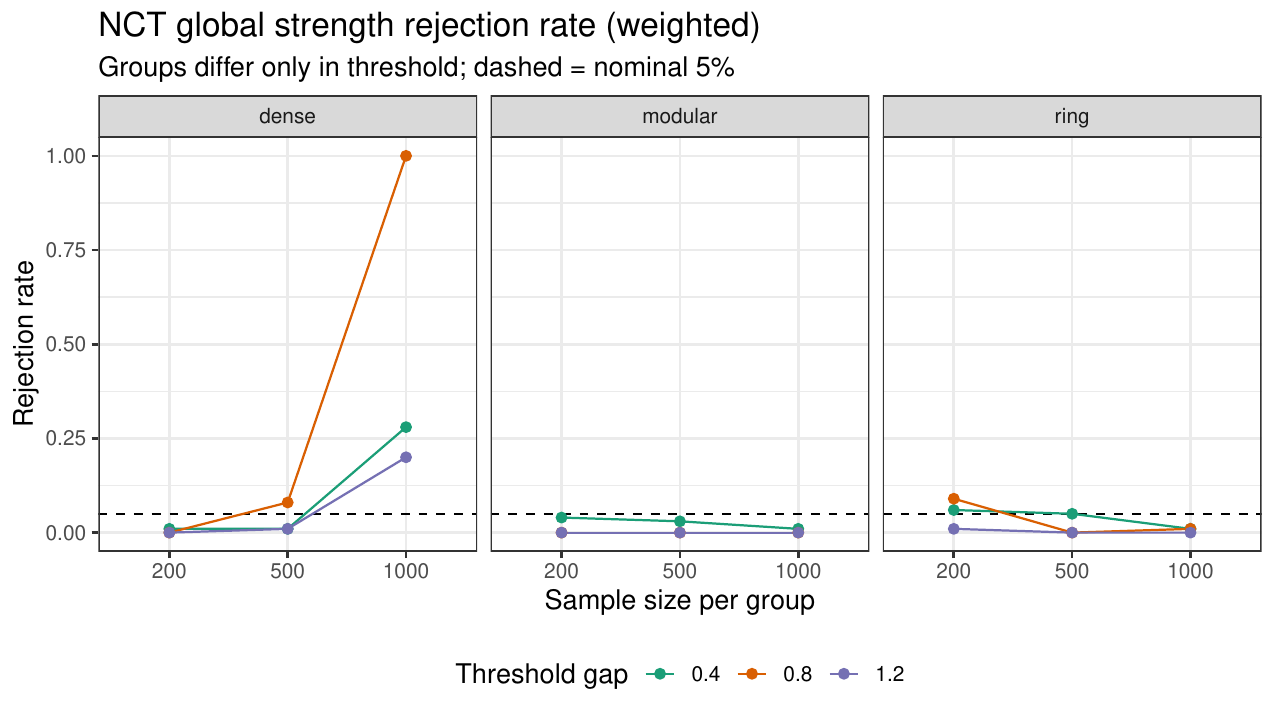}
  \caption{\textbf{Supplementary Monte Carlo check of NCT under threshold differences only.}
  Two groups were simulated with identical coupling but different thresholds and compared using weighted NCT. The dashed horizontal line marks the nominal 5\% level under the null hypothesis that the two groups have the same coupling. Shown here are rejection rates for the structure invariance test and the global strength test.}
  \label{fig:appendix_nct_part1}
\end{figure}

\begin{figure}[htbp]
  \centering
  \includegraphics[width=\linewidth]{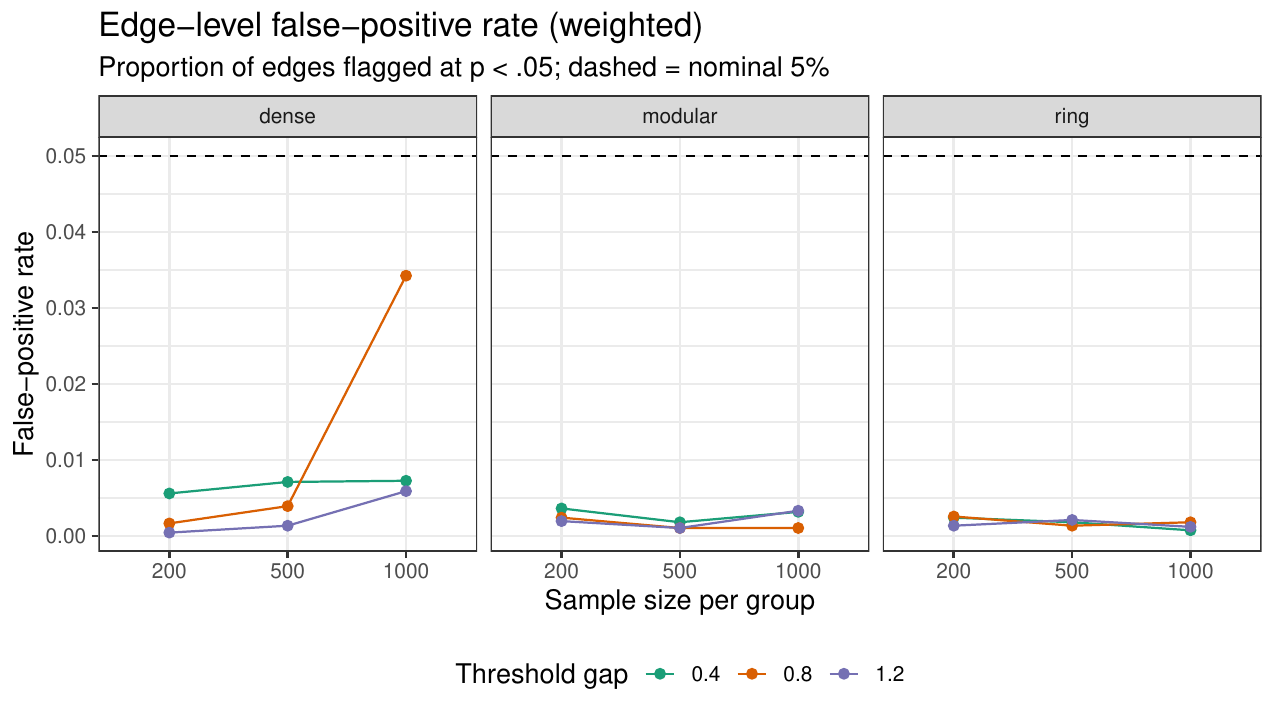}
  \caption{\textbf{Supplementary Monte Carlo check of NCT under threshold differences only: edge-level tests.}
  Edge-level false-positive rates from the same simulation. The dashed horizontal line marks the nominal 5\% level. False-positive rates remained low overall.}
  \label{fig:appendix_nct_part2}
\end{figure}

The results were mixed. The structure invariance test remained broadly close to
the nominal 5\% level, and edge-level false-positive rates were low overall. By
contrast, the global strength test showed stronger inflation in dense networks,
especially at the largest sample size. Thus, this supplementary analysis
suggests that threshold differences do not necessarily create broad false
evidence of different network structure, but they can still generate misleading
differences in estimated global strength. Figures~\ref{fig:appendix_nct_part1}
and~\ref{fig:appendix_nct_part2} summarize the rejection rates and edge-level
false-positive rates.

\end{document}